\documentclass[]{iopart}

\expandafter\let\csname equation*\endcsname\relax
\expandafter\let\csname endequation*\endcsname\relax

\usepackage[unicode]{hyperref}
\usepackage[all]{hypcap}
\usepackage{amssymb,amsmath,verbatim,mathtools,needspace,enumitem,etoolbox,graphicx,microtype,pifont,url,cite}
\usepackage[usenames,dvipsnames]{xcolor}
\usepackage[T1]{fontenc}
\usepackage[utf8]{inputenc}
\usepackage[normalem]{ulem}
\definecolor{linkcolor}{rgb}{0.0,0.3,0.5}
\hypersetup{colorlinks=true,linkcolor=linkcolor,citecolor=linkcolor,filecolor=linkcolor,urlcolor=linkcolor}

\newcommand{\nn}{\nonumber}

\newcommand{\milan}{{Dipartimento di Fisica ``G. Occhialini'', Universit\'a degli Studi di Milano-Bicocca, Piazza della Scienza 3, 20126 Milano, Italy}}
\newcommand{\infn}{{INFN, Sezione di Milano-Bicocca, Piazza della Scienza 3, 20126 Milano, Italy}}
\newcommand{\duke}{{Department of Physics, Duke University, 120 Science Drive, Durham, NC 27708, USA}}

\newcommand{\baylor}{{Department of Physics and Astronomy, Baylor University, One Bear Place 97316, Waco, TX 76798-7316, USA}}

\allowdisplaybreaks

\begin{document}

\begin{center}
\title[N.~Loutrel et al.]{Probing modified gravitational-wave dispersion\\with bursts from eccentric black-hole binaries}
\end{center}

\author{
Nicholas Loutrel$^{1,2}$, %\orcidlink{0000-0002-1597-3281}, 
Ava Bailey$^{3,4}$, %\orcidlink{0009-0002-0639-906X},
Davide Gerosa$^{1,2}$, %\orcidlink{0000-0002-0933-3579}
}
\vspace{0.1cm}
\address{$^{1}$~\milan}
\address{$^{2}$~\infn}
\address{$^{3}$~\duke}
\address{$^{4}$~\baylor}

\ead{\href{mailto:nicholas.loutrel@unimib.it}{\rm nicholas.loutrel@unimib.it}}

\setcounter{footnote}{0}

\begin{abstract}
Gravitational waves in general relativity are non-dispersive, yet a host of modified theories predict dispersion effects during propagation. In this work, we consider the impact of dispersion effects on gravitational-wave bursts from highly eccentric binary black holes. We consider the dispersion effects within the low-energy, effective field theory limit, and model the dispersion relation via standard parameterized deformations. Such modified dispersion relations produce two modifications to the burst waveform: a modification to the time of arrival of the bursts in the detector, which appears as a 2.5PN correction to the difference in burst arrival times, and a modification to the arrival time of individual orbital harmonics within the bursts themselves, resulting in a Bessel-type amplitude modulation of the waveform. Using the Fisher information matrix, we study projected constraints one might obtain with future observations of repeating burst signals with LIGO. We find that the projected constraints vary significantly depending on the theoretical mechanism producing the modified dispersion. For massive gravitons and multifractional spacetimes that break Lorentz invariance, bounds on the coupling parameters are generally weaker than current bounds. For other Lorentz invariance breaking models such as Ho\v{r}ava-Lifschitz gravity, as well as scenarios with extra dimensions, the bounds in optimal cases can be 1-3 orders of magnitude stronger than current bounds. 
\end{abstract}

\section{Introduction}
\label{sec:intro}
%%%%%%%%%%%%%%%%%%%%%%%%%%%%%%%%%%%%%%%%%%%%%%%%%%%%%%%%%%%%%%%%%%%%%%%%%%%%%%%%%%%%%%%%%%%%%%%%%%%%%

In general relativity (GR), gravitational waves (GWs) travel at the speed of light regardless of their wavelength. Thus, GWs in GR are said to be non-dispersive, a fact that is confirmed to high confidence by the third GW transient catalog (GWTC-3)~\cite{LIGOScientific:2021sio}, as well as the combined observations of electromagnetic emission and GWs from binary neutron star mergers~\cite{Baker:2017hug,LIGOScientific:2018dkp}. 

However, from a mathematical standpoint, GR is known to possess unresolved theoretical issues~\cite{Ashtekar:2022oyq,Carlip:2001wq}. In particular, the theory is not UV complete, leading to an incompatibility with quantum physics~\cite{Carlip:2001wq, Gubitosi:2019ncx, Eichhorn:2018yfc, Sen:2021ffc, PhysRevD.97.086006, universe8040229}. Many attempts have been made to resolve this conflict, a few among many being string theory~\cite{Green:2012oqa,Green:2012pqa,1998stth.book.....P}, loop quantum gravity~\cite{Ashtekar:2021kfp,2007mcqg.book.....T}, causal dynamical triangulations~\cite{Loll:2019rdj, Ambjorn:2024pyv}, and Ho\v{r}ava-Lifschitz gravity~\cite{Horava:2008ih,Horava:2009uw}. All of these high energy theories predict new physics appearing at the relevant energy scale associated with the theory. More specifically, string theory predicts the existence of additional spacetime dimensions, while quantum gravity theories often break Lorentz invariance through the introduction of a minimum length scale. Generally, these energy and length scales are taken to be the Planck scale for theoretical reasons, with experimental particle physics results indicating that these scales cannot be less than the electroweak scale~\cite{Hoyle:2004cw,MICROSCOPE:2022doy,CMS:2020kwy}.

Probing such high-energy theories is a key, open challenge in modern physics. While a complete UV theory of gravity is currently not known, even in the low‐energy, so‐called effective field theory (EFT) limit, modifications to GR are expected, and these effects are often used to construct phenomenological models of beyond‐GR physics~\cite{Levi:2018nxp,Alexander:2009tp,Donoghue:1994dn,Donoghue:2012zc,Donoghue:2015hwa,Donoghue:2022eay,Capozziello:2013ava,Burgess:2003jk}. Such models are parameterized by, effectively, undetermined coupling parameters that capture the leading order modifications to GR and can then be constrained by current experiments. This is the central idea behind the parameterized post-Einsteinian formalism (PPE)~\cite{Yunes:2009ke,Loutrel:2014vja, Loutrel:2022xok, Tahura:2018zuq} targeting tests of GR with GWs, in analogy to the parameterized post-Newtonian (PPN)~\cite{Will:2014kxa,Hohmann:2021gbq} and post-Keplerian (PPK)~\cite{1985AIHPA..43..107D,1986AIHPA..44..263D,Damour:1991rd} formalisms for Solar System and binary pulsar tests, respectively. The ppE formalism complements current tests of GR with GWs~\cite{LIGOScientific:2021sio, Gupta:2020lxa, Mahapatra:2023uwd, Mahapatra:2025cwk} and has provided the link to mapping theory agnostic constraints on the amplitudes of GW phase deformations to specific theories~\cite{Yunes:2016jcc,Carson:2020rea}. 

The effects the ppE formalism seeks to constrain can be split into two sets: generation effects, which modify how the GWs are generated by a binary, and propagation effects, which modify the propagation of GWs as they travel from source to detector. Dispersion, which considers GWs propagating at different speeds, falls into the latter category. This effect is modeled by modifying the standard GR dispersion relation for GWs, and allowing it to become momentum (or wavelength, depending on how the dispersion relation is written) dependent. These modifications to the dispersion relation specifically capture the leading order modifications to GW propagation in GR, despite a complete UV theory not current being known~\cite{Will:1997bb}. For an extensive analysis of how modified dispersion impacts GWs from quasi-circular binaries, see Refs.~\cite{Mirshekari:2011yq,Samajdar:2017mka}.

A great deal of work in population synthesis has, however, shown that not all GWs detected by LIGO are expected to originate from quasi-circular binaries. There is even evidence that some of the detections in GWTC-3 may possess non-negligible eccentricity~\cite{Romero-Shaw:2022xko,Romero-Shaw:2025vbc,Gupte:2024jfe}, including GW190521~\cite{LIGOScientific:2020iuh}, which  may have been generated from a high-eccentricity merger~\cite{Gamba:2021gap}. Although such high eccentricity events are rare~\cite{Rodriguez:2018pss, Tagawa:2020jnc}, both modeled~\cite{Nagar:2020xsk} and unmodeled~\cite{Tai:2014bfa} search strategies currently exist to detect such signals, and they may prove to be important laboratories for probes of fundamental physics. Tests of GR with GWs from eccentric binaries, much like waveform modeling efforts, have historically lagged behind those of quasi-circular binaries. However, a few analyses for low and moderate eccentricity binaries have shown that orbital eccentricity can have a non-monotonic impact on bounds on beyond-GR physics~\cite{Ma:2019rei,Moore:2020rva}. Further, neglecting eccentricity in waveform modeling can lead to false claims of new physics~\cite{Gupta:2024gun,Saini:2022igm,Bhat:2022amc,Narayan:2023vhm,Saini:2023rto}.
%through model comparison, it has been shown that beyond-GR physics can be confused with effects due to orbital eccentricity, and vice versa.

In this work, we seek to fill in one of these gaps by considering probes of modified GW dispersion with eccentric binaries, focusing on highly eccentric burst sources with ground-based GW detectors. We study the high eccentricity limit for one reason, specifically, highly eccentric binaries emit GWs at every overtone of the orbital frequency, as opposed to quasi-circular and low eccentricity sources. In the time domain, the GW signal resembles a repeating burst signal, with bursts of radiation emitted at each pericenter passage. Meanwhile in the frequency domain, the GWs are broadband, and peak at a harmonic number that can be significantly greater than unity~\cite{Wen:2002km}. This broadband emission makes them ideal candidates for considering modified dispersion effects where GWs can travel at different velocities depeonding on their wavelength or frequency.

To this end, we employ the effective fly-by (EFB) formalism we developed in Refs.~\cite{Loutrel:2019kky,Loutrel:2020jfx}, with some slight modifications to the Fourier domain waveform that are detailed in Appendix~\ref{app:efb-new} herein. When including modified dispersion effects into the EFB waveform model, two modifications appear. The first is an amplitude modulation, due to the fact that different orbital harmonics within the burst travel at different velocities. The modulation is captured in both the time and frequency domains by a Bessel-type generating functions, which unfortunately does not cleanly fit into the standard PPE formalism.

The second modification appears in the time of arrival of the bursts in the detector frame. In GR, the burst can be modeled as having a time-frequency centroid $(t_{i}, f_{i})$ where the time between bursts $\Delta t_{i,i-1} = t_{i} - t_{i-1}$ is related to the parameters of the binary through the orbital period and radiation reaction effects. Including modified dispersion into this creates a new 2.5PN deformation that captures all dispersion effects that fit into the standard EFT dispersion relation given by Eq.~\eqref{eq:disp-relation} below. Putting both of these effects together, we obtain an EFB waveform model with parameterized dispersion effects, that is both purely analytic and fast to evaluate.

Since the waveform model is analytic, we study what bounds can be placed on dispersion effects with observations of highly eccentric binaries using the Fisher information method~\cite{Finn:1992wt,Cutler:1994ys}, which is commonly used in GW forecasting (see~\cite{Vallisneri:2007ev} for some limitations). We consider a small subset of such binaries as sources for a single LIGO GW detector. How stringent these bounds are depends heavily on the type of dispersion effect considered. More specifically, it depends strongly on the exponent parameter $a$ that controls the power law modification to the GW dispersion relation. The larger $a$, the better constraints obtained. For dispersion effects with $a<3$, such as gravitons possessing a mass, we generally find that the projected constraints do not beat current bounds from GWTC-3. On the other hand, for effects with $a\ge 3$, the bounds from single eccentric burst sources can beat the current GWTC-3 bounds by up to six orders of magnitude for specific theoretical scenarios, and under optimal binary parameters. More specifically, binaries with moderately high initial eccentricity ($e \sim 0.8$) and higher chirp masses (at fixed luminosity distance) lead to better constraints on modified dispersion effects. 

This paper is organized as follows. In Sec.~\ref{sec:primer}, we present a brief overview of dispersion effects in the EFT limit, and provide details of some example theories. In Secs.~\ref{sec:time-domain} and~\ref{sec:efb}, we derive the modifications to the EFB burst waveforms in the time and frequency domain, respectively. The modification to the burst arrival times is derived in Sec.~\ref{sec:timing}, while Sec.~\ref{sec:fisher} presents the details for our Fisher analysis. Finally, the bounds obtained from the Fisher analysis are presented in Sec.~\ref{sec:constraints}, and we provide discussion in Sec.~\ref{sec:disc}. 
%The main results of this work are summarized in Fig.~\ref{fig:violins} \dg{if we refer to a figure here, that cannot be Fig 6 but become Fig 1, and should also be placed here (I tried before, the PRD editors will change this). I would either describe the figure here (executive summary?) or reprahse this to remove the figure/table references} and Table~\ref{tab:summary}, which includes a mapping from GWTC-3 theory agnostic bounds on dispersion to the bounds on theory specific parameters that, to our knowledge, as not been published before. 
We work in geometrized units where $G = c = 1$. Note that in this unit system, for Planck's constant one has $\hbar = L_{P}^{2}$ where $L_{P}$ is the Planck length.

\begin{table*}[t]
    \centering
    \resizebox{\textwidth}{!}{%
    \begin{tabular}{c|c|c|c|c|c}
      % \hline\hline %\vspace{-0.2em}
         & & Theory & & Current & Projected\\
         Theoretical mechanism & GR pillar & parameter & $a$ & GW bound~\cite{LIGOScientific:2021sio} & EGWB bound \\
         \hline\hline
         Massive gravity & $m_{g} = 0$ & $m_{g}$ [eV] & $0$ & $1.27 \times 10^{-23}$ & $4.8\times10^{-23}$\\
         \hline
         Extra dimensions & 4D & $\alpha_{\rm edt}$ & $4$ & $1.10\times 10^{60}$ & $5.7\times10^{58}$ \\
         \hline
         Doubly special relativity & LI & $\eta_{\rm drst}$ & $3$ & $2.84 \times 10^{20}$ & $1.3\times10^{19}$ \\
         \hline
         Ho\v{r}ava-Lifschitz gravity & LI & $\kappa_{hl}^{4}\mu_{hl}^{2}$ [eV$^{-2}$] & $4$ & $0.74 \times 10^{4}$ & $0.039\times10^{4}$ \\
         \hline
         Multifractional spacetime & LI & \begin{tabular}{@{}c@{}} $\omega_{\star}$ [eV$^{-1}$] (time) \\ $\omega_{\star}$ [eV$^{-1}$] (space) \end{tabular} & $\in[2,3]$ & \begin{tabular}{@{}c@{}} $5.44\times 10^{-29}(a=2.5)$ \\ $2.27\times 10^{-28} (a=2.5)$\end{tabular} & \begin{tabular}{@{}c@{}} $1.3\times10^{-29}$ \\ $ 4.9\times10^{-31}$ \end{tabular} \\
         \hline
         Gravitational SME & LI & \begin{tabular}{@{}c@{}} $k^{d}_{(I)}$ [cm$^{d-4}$] (even) \\ $+k^{(d)}_{(V)}$ [cm$^{d-4}$] (odd) \\ $-k^{(d)}_{(V)}$ [cm$^{d-4}$] (odd) \end{tabular} & $d-2$ & \begin{tabular}{@{}c@{}} $1.17\times10^{-2} (d=6)$ \\ $1.30\times10^{-11} (d=5)$ \\  $4.60\times10^{-11} (d=5)$ \end{tabular} & \begin{tabular}{@{}c@{}} $1.5\times10^{-7}$ \\ $2.0\times10^{-14}$ \\ $2.0\times10^{-14}$ \end{tabular}\\
%         \hline \hline
    \end{tabular}
    }
    \caption{Table summarizing the various theoretical mechanisms (first column) that lead to modified GW dispersion. Each of these theoretical mechanisms breaks a fundamental pillar of GR (second column), where LI is short for Lorentz invariance. The third column provide the coupling parameter that appears in $\alpha$ in Eq.~\eqref{eq:disp-relation}, while the fourth column provides the exponent parameter. The fifth column shows the current bounds on the coupling parameters found by mapping the theory agnostic constraints from GWTC-3 (Table VII of~\cite{LIGOScientific:2021sio}) to each specific scenario. Lastly, the final column show the projected bounds obtained in the work from eccentric GW burst (EGWB) source. See Sec.~\ref{sec:params} for the full details of how these bounds are obtained.}
    \label{tab:summary}
\end{table*}

%%%%%%%%%%%%%%%%%%%%%%%%%%%%%%%%%%%%%%%%%%%%%%%%%%%%%%%%%%%%%%%%%%%%%%%%%%%%%%%%%%%%%%%%%%%%%%%%%%%%%
\section{Modified Dispersion in Eccentric Binaries}
\label{sec:theory}
%%%%%%%%%%%%%%%%%%%%%%%%%%%%%%%%%%%%%%%%%%%%%%%%%%%%%%%%%%%%%%%%%%%%%%%%%%%%%%%%%%%%%%%%%%%%%%%%%%%%%

%-----------------------------------------------
\subsection{Modified gravitational-wave dispersion: a primer}
\label{sec:primer}
%-----------------------------------------------

A host of different non-GR phenomena can modify the propagation of GWs, but only a subset of these result in dispersion. Dispersion effects may be captured in the GWs through the parameterized dispersion relation~\cite{Yunes:2016jcc,Carson:2020rea,Mirshekari:2011yq}
\begin{equation}
    \label{eq:disp-relation}
    \omega^{2} = k^{2} + \alpha L_{P}^{2a-4} k^{a},
\end{equation}
where $(\omega, k)$ are the angular frequency and wave number of the graviton, and $(\alpha, a)$ are non-GR parameters that capture the strength and type of dispersion effect, respectively. The above parameterized dispersion relation causes GWs at different frequencies to propagate at different velocities, which can be seen from the group velocity of the gravitons, specifically
\begin{equation}
    v_{g} = \frac{d\omega}{dk} = 1 + \frac{(a - 1)}{2} \alpha  L_{P}^{2a-4} \omega^{a-2}\,.
\end{equation}
Due to the modification of the above group velocity, gravitons emitted at the same instant, but at different frequencies, will arrive at a GW detector at different times. A thorough calculation of this effect, including the influence of cosmological redshift, can be found in Ref.~\cite{Mirshekari:2011yq}. For simplicity, we neglect the effect of cosmological backgrounds\footnote{There is a further, more physical motivation to neglect cosmological backgrounds in this study. Individual bursts from highly eccentric binaries are generally not very loud and are expected to have modest SNRs (see~\cite{Loutrel:2020jfx}). As such, they are unlikely to be detectable at large cosmological distances with LIGO, where cosmological effects become important.
}, and thus, for two gravitons of frequency $(f,f')$ and emitted at times $(t_e, t_e')$, the observed times of arrival are related by
\begin{equation}
    \label{eq:observed-time}
    \Delta t_{o} = \Delta t_{e} + \frac{D_{L}}{2 \lambda_{\alpha}^{2-a}} \left(\frac{1}{f^{2-a}} - \frac{1}{f'^{2-a}}\right), 
\end{equation}
where $D_{L}$ is the (luminosity) distance to the source, $\Delta t_{o,e} = t_{o,e} - t_{o,e}'$, and
\begin{equation}
    \lambda_{\alpha} = 2\pi L_{P}^{2} \alpha^{1/(a-2)}\,,
\end{equation}
which is the ``wavelength'' associated with $\alpha$.

The parameters $(\alpha, a)$ map to specific non-GR scenarios. Many of these arise due to the introduction of a minimal length scale and the breaking of Lorentz invariance, since these concepts are usually connected. A few examples include:
\begin{itemize}
    \item \textit{Massive gravitons}~\cite{Will:1997bb,Rubakov:2008nh,Hinterbichler:2011tt,deRham:2014zqa}: The simplest, and largely phenomenological, deviation is for gravitons to possess a mass $m_{g}$. In this case, the dispersion relation obtains a momentum/wave number independent term. Thus $\alpha = m_{g}^{2}$ and $a = 0$, and hence $\lambda_{g} = 2\pi L_{P}^{2}/m_{g}$ is the Compton wavelength of the graviton.
    \item \textit{Extra dimensions}~\cite{Sefiedgar:2010we}: A postulated dispersion relation obtained by comparing to a generalized uncertainty principle in a universe with extra dimensions reveals $\alpha = \alpha_{\rm edt}/L_{P}^{2}$ and $a=4$, where $\alpha_{\rm edt}$ is a dimensionless constant. In this case $\lambda_{\rm edt} = 2\pi L_{P} \sqrt{\alpha_{\rm edt}}$.
    \item \textit{Doubly special relativity}~\cite{Amelino-Camelia:2000cpa,Magueijo:2001cr,Amelino-Camelia:2002cqb}: A modification of special relativity through nonlinear modification of the Lorentz group by introducing an invariant length/momentum scale, in addition to the invariant speed of light. The length scale is taken to be the Planck length, and in the limit of the wavelength of the wave being much larger than this scale, $\alpha = \eta_{\rm drst}/L_{P}$ and $a = 3$, where $\eta_{\rm drst}$ is here dimensionless. In this case $\lambda_{\rm drst} = 2\pi L_{P} \eta_{\rm drst}$.
    \item \textit{Ho\v{r}ava-Lifschitz gravity}~\cite{Horava:2008ih,Horava:2009uw,Vacaru:2012nvq,Blas:2011zd}: A proposed quantum theory of gravity that introduces a preferred time foliation, such that Lorentz invariance only arises at large distances compared to the Planck length. In this theory, GWs obey a dispersion relation with $\alpha= \kappa_{hl}^{4} \mu_{hl}^{2}/16$ and $a = 4$, where $(\kappa_{\rm hl}, \mu_{\rm hl})$ are coupling constants of the theory related to the bare gravitational constant and balance conditions of the theory, respectively.
    \item \textit{Multifractional spacetime theory}~\cite{Calcagni:2009kc,Calcagni:2011kn,Calcagni:2011sz,Calcagni:2016zqv}: Another proposed quantum theory of gravity that allows the effective dimension of spacetime to vary at different scales by replacing the $D$-dimensional measure of the action $d^{D}x$ with a Lebesgue-Stieltjes measure $d\rho(x)$. For this theory, the GW dispersion relation takes different forms depending on the preferred foliation of the spacetime. For timelike fractal foliations, $\alpha = 2 \omega_{\star}^{2-a}/(3-a)$, while for spacelike fractal foliations $\alpha = -2\times3^{1-a/2}\omega_{\star}^{2-a}/(3-a)$, with $\omega_{\star}$ a characteristic frequency (energy) scale. For both scenarios, $a$ need not be an integer values, but $a=2-3$ and typically one choses $a=2.5$.
    \item \textit{Gravitational standard model extension (SME)}~\cite{Kostelecky:2016kfm}: An effective field theory that combines the standard model of particle physics with GR, and includes Lorentz symmetry breaking. The exact form of the modified dispersion relation varies depending on the dimension of the Lorentz breaking operators introduced into the action. For even $d\ge 4$, $\alpha = -2k_{(I)}^{(d)}/L_{P}^{2d-8}$, while for odd $d\ge 5$, $\alpha = \pm k_{(V)}^{(d)}/L_{P}^{2d-8}$, and $a=d-2$. The constant coefficients $k_{(I,V)}^{(d)}$ control the Lorentz violating operators in the action. Note that each $k^{(d)}$ has units of length$^{d-4}$.
\end{itemize}
The above list is far from exhaustive, but provides a general sense of the physics that the modified dispersion relation in Eq.~\eqref{eq:disp-relation} captures. For further details of these scenarios, see Ref.~\cite{Yunes:2016jcc} and references therein. In the next section, we will visually show the imprint of a few of these cases on the propagation of eccentric bursts.
%

%-----------------------------------------------
\subsection{Eccentric binaries and time-domain gravitational waves}
\label{sec:time-domain}
%-----------------------------------------------

The first formulation of the gravitational waveform for eccentric binaries at leading post-Newtonian (PN)~\cite{Blanchet:2013haa} order was derived by Wahlquist \cite{Wahlquist:1987rx} and later reformed with newer notation by Martel and Poisson \cite{Martel:1999tm}. We follow Martel and Poisson and define the GW polarizations as
\begin{equation}
\begin{split}
    h_{+} = & -\frac{m \eta}{p D_{L}} \Bigl\{\Bigl[2\cos(2\phi - 2\beta) + \frac{5}{2}e\cos(\phi - 2\beta)  \\ 
    & + \frac{1}{2}e\cos(3\phi-2\beta) + e^{2}\cos2\beta\Bigr](1 + \cos^{2}\iota)  \\ & + [e\cos\phi + e^{2}]\sin^{2}\iota\Bigr\},
\end{split}
\label{eq:hplus}
\end{equation}
and
\begin{equation}
\begin{split}
    h_{\times} = & -\frac{m \eta}{p D_{L}} [4\sin(2\phi - 2\beta) + 5e\sin(\phi - 2\beta)  \\ & + e\sin(3\phi-2\beta) - 2e^{2}\sin2\beta]\cos\iota,
\end{split}
\label{eq:hcross}
\end{equation}
where $\beta$ and $\iota$ are the angles that define the polarization axes, $\eta$ is the symmetric mass ratio, $m$ is the total mass of the system, $e$ is the orbital eccentricity, $p$ is the semilatus rectum which is related to eccentricity and semi-major axis by $p = a (1-e^{2})$, and $D_{L}$ is the luminosity distance.

The time-domain signal of the waveforms can be decomposed into harmonics of the mean orbital frequency~\cite{Moreno-Garrido:1995sxt,Moore:2018kvz}
%which shows the time dependence of the field component equations
\begin{equation}
    h_{+,\times}(t) = -h_{0}\sum_{k=1}^{\infty}\Bigl[S_{+,\times}^{(k)}\sin(k\ell) + C_{+,\times}^{(k)}\cos(k\ell)\Bigr],
\label{eq:time_polarizations}
\end{equation}
where we introduce the notation used in Ref.~\cite{Moore:2018kvz} such that the coefficients $S_{+,\times}^{(k)}$ and $C_{+,\times}^{(k)}$ are functions of eccentricity and the polarization angles and $h_{0}$ is dependent upon the individual masses of the binary constituents, semi-major axis, eccentricity, and luminosity distance. These coefficients are explicitly defined as:
\begin{align}
    \label{eq:Spk}
    S_{+}^{(k)} &= \frac{4}{e^{2}} [1+\cos^{2}\iota] \sin 2\beta \sqrt{1-e^{2}} \Bigl\{e J_{k-1}(ke) 
    %\nn\\ &
    - [1+k(1-e^{2})] J_{k}(ke)\Bigr\},
    \\
    C_{+}^{(k)} &= \frac{2}{e^{2}} \Bigl\{ \Bigl[\cos2\beta (1+\cos^{2}\iota) e (1-e^{2}) k \bigl(J_{k+1}(ke) 
    %\\ & 
    - J_{k-1}(ke)\bigr)\Bigr] 
    \nn\\ & - 
        \Bigl[\cos2\beta(1+\cos^{2}\iota)(e^{2}-2) 
        %\nn\\ & 
        + e^{2}\sin^{2}\iota\Bigr]J_{k}(ke)\Bigr\},
    \\
    S_{\times}^{(k)} &= \frac{8}{e^{2}} \cos2\beta \cos\iota \sqrt{1-e^{2}} \Bigl\{e J_{k-1}(ke) 
    %\nn\\ &
     - [1 + (1-e^{2})k] J_{k}(ke)\Bigr\},
    \\
    \label{eq:Ckc}
    C_{\times}^{(k)} &= \frac{4}{e^{2}} \sin2\beta \cos\iota \Bigl\{2e(1-e^{2})k J_{k-1}(ke)  
    %\nn\\ &
     -2\Bigl[1+(1+e^{2})k-\frac{e^{2}}{2}\Bigr]J_{k}(ke)\Bigr\},
    \\
    h_{0} &= \frac{2m_{1}m_{2}}{a(1-e^{2})} \frac{2}{D_{L}}.
\end{align}
Equation~\eqref{eq:time_polarizations} is also dependent upon the mean anomaly $\ell$ which is given as $\ell = 2 \pi (t-t_{p}) F_{\rm orb}$, with $F_{\rm orb}$ being the orbital frequency, and $t_{p}$ the time of pericenter passage. 

These waveforms describe a continuous radiative signal that is non-dispersive. Under modifications to GR, the waveforms become dispersive requiring the study and consideration of time shifts between the time at which GWs are emitted and when they are observed. At leading PN order, quasi-circular binaries only emit GWs at twice the orbital frequency, which can be seen by taking the limit $e\rightarrow0$ of Eqs.~\eqref{eq:hplus}-\eqref{eq:hcross}. As a result, dispersion effects modify the the observed frequency evolution of the binary (see Ref.~\cite{Mirshekari:2011yq} for details). However, comparison to the eccentric case in Eq.~\eqref{eq:time_polarizations} reveals that eccentric binaries emit GWs in many orbital harmonics simultaneously. Since each harmonic possesses a different frequency, they will arrive at the detector at different times, resulting in an observed modulation of the waveform. Here, we are primarily interested in the high eccentricity limit, where the waveforms resemble discrete bursts instead of continuous waves. While the discussion will primarily focus on this case, the formalism we develop here is general enough to apply to low and moderately eccentric systems as well.
%{\color{red}\sout{This manifests as a modification to the mean anomaly such that the time-dependent field component expressions no longer depend on $t$ which, in the aforementioned case, is synonymous with both the time at which the GW is emitted and the time of observation but now depend on a frequency-dependent time shift between the GW's emitted and observed times.}}

%In our initial discussion of mean anomaly $\ell$ as is included in
The mean anomaly $\ell$ appearing in Eq.~\eqref{eq:time_polarizations}
%we highlighted that $\ell$ 
is a function of coordinate time $t$ which, in GR, is measured without distinction between emitted and observed times. Under dispersion effects and modifications to GR, the time $t$ which characterizes the mean anomaly maps to emitted time $t_{e}$, i.e. the coordinate time at which the GW is emitted in the binary's center-of-mass rest frame.
%considering this is the absolute time at which the GW is radiated as measured in respect to the system and the system's time of pericenter passage. 
Under dispersion, the mean anomaly should be expressed as $\ell_{e} = 2 \pi F_{\rm orb} (t_{e} - t_{p}^{(e)}) = 2\pi F_{\rm orb} \Delta t_{e}$, where $t_{p}^{(e)}$ is the ``emitted'' time of pericenter passage, and we relabel $\ell$ with the `$e$' subscript to indicate its dependence on the emitted time. Each harmonic in Eq.~\eqref{eq:time_polarizations} will propagate at a different velocity, and will thus be observed at time $t_{o}$. We further define $\ell_{o} = 2\pi F_{\rm orb}(t_{o} - t_{p}^{(o)}) = 2\pi F_{\rm orb} \Delta t_{o}$, with $t_{p}^{(o)}$ the ``observed'' time of pericenter passage. For the $k$-th harmonic of the waveform, the quantities $\ell_{o}$ and $\ell_{e}$ are then related to each other by Eq.~\eqref{eq:observed-time} taking $f = k F_{\rm orb}$,
\begin{equation}
    \label{eq:l_o}
    \ell_{o} = \ell_{e} + \frac{\pi D_{L}}{\lambda_{\alpha}^{2-a} F_{\rm orb}^{1 - a}} \left(\frac{1}{k^{2-a}} - \frac{1}{k_{\rm max}^{2-a}} \right)\,,
\end{equation}
where we have also chosen $f'=k_{\rm max} F_{\rm orb}$, with~\cite{Wen:2002km}
\begin{equation}
    k_{\rm max} = \frac{2 (1+e)^{1.1954}}{(1-e^{2})^{3/2}}\,,
\label{eq:kmax}
\end{equation}
which corresponds to the harmonic with maximum power.

We have made two choices to arrive at Eq.~\eqref{eq:l_o}, namely that the constant time shifts physically associated with pericenter passage $[t_{p}^{(e)}, t_{p}^{(o)}]$ are different for $[\ell_{e}, \ell_{o}]$, and that the reference frequency $f'$ is chosen to be fixed at $k_{\rm max} F_{\rm orb}$. Both of these stem from the burst formalism originally developed for power stacking searches in Ref.~\cite{Loutrel:2014vja}. There, eccentric bursts were treated as regions of excess power in a time-frequency spectrogram, each of which possessing a central coordinate $(t_{i}, f_{i})$. Since the modelling therein was performed in the context of GR (i.e. no dispersion), the time centroid is trivially the time of pericenter passage of the orbit from which the burst originates, while the frequency centroid is related to the peak of the waveform power spectrum or amplitude. The former can be mapped to the pericenter passage time of the previous orbit by incorporating radiation reaction effects into a timing model, which we will discuss in more detail in Sec.~\ref{sec:timing}. The latter was first considered by Turner~\cite{1977ApJ...216..610T}, who showed that the peak of the power spectrum for nearly parabolic orbits is related to the characteristic timescale of pericenter passage, although a more thorough computation was carried out in Ref.~\cite{Wen:2002km}, leading to Eq.~\eqref{eq:kmax}. Making these choices will allow us to address how the time-frequency centroid of each burst is modified due to dispersion effects in Sec.~\ref{sec:timing}.

At this point, we may implement the dispersion effects into the time-domain waveform.
%becomes useful to consider implementing Bessel modulations in the field component equations. 
We are able to express Eq.~\eqref{eq:time_polarizations} in terms of complex exponentials with an harmonic coefficients dependent upon $S_{+,\times}^{(k)}$ and $C_{+,\times}^{(k)}$ through manipulating Euler's formula. This gives a new, equivalent expression for the polarizations,
\begin{equation}
    h_{+,\times}(t_{e}) = -h_{0} \sum_{k=1}^{\infty} \Bigl[H^{(k)}_{+,\times} e^{ik\ell_{e}} + \left(H^{(k)}_{+,\times}\right)^{\dagger}e^{-ik\ell_{e}}\Bigr],
\label{eq:BD}
\end{equation}
where $H_{+,\times}^{(k)} = (C_{+,\times}^{(k)} - iS_{+,\times}^{(k)})/2$, and $\dagger$ indicates the complex conjugate of the specified terms. Now that the time-dependent term $\ell_{e}$ has been isolated into an exponential term, we use Eq.~\eqref{eq:l_o} to separate the exponential into a product of two terms one of which depends solely on $\ell_{o}$ and another which is time-independent, specifically
\begin{equation}
\begin{split}
    %e^{ikl_{e}} & = e^{ikl_{o}} \cdot e^{-ikl}\\ & = e^{ik2\pi t_{o} F_{\rm orb}} \cdot e^{-ik2 \pi F_{\rm orb} \alpha \bigl(\frac{1}{F_{\rm orb}^{2} k^{2}} - \frac{1}{F_{\rm orb}^{2} k_{\rm max}^{2}}\bigr)}
    e^{ik \ell_{e}} = B_{k}(\bar{\alpha}) e^{ik\ell_{o}}\,,
\label{eq:BDexp}
\end{split}
\end{equation}
where
\begin{equation}
    \label{eq:B_k}
    B_{k}(\bar{\alpha}) = \exp\left\{\frac{i}{2} \bar{\alpha} k^{a/2} \left[ \left(\frac{k}{k_{\rm max}} \right)^{1-\frac{a}{2}} - \left(\frac{k}{k_{\rm max}} \right)^{\frac{a}{2} -1}\right] \right\}\,,    
\end{equation}
with the dimensionless coupling parameter
\begin{equation}
    \label{eq:alpha_bar}
    \bar{\alpha} = \frac{2\pi D_{L}k_{\rm max}^{(a/2)-1}}{\lambda_{\alpha}^{2-a} F_{\rm orb}^{1-a}}\,.
\end{equation} 
Thus, the final expression of the dispersion modulated waveform is 
\begin{equation}
\begin{split}
    h_{+,\times}(t_{o}) = & -h_{0} \sum_{k=1}^{\infty} \Bigl[ H^{(k)}_{+,\times} B_{k}(\bar{\alpha}) e^{i k \ell_{o}} 
    %\\& 
    + \left(H^{(k)}_{
    +,\times}\right)^{\dagger} B^{\dagger}_{k}(\bar{\alpha}) e^{-i k \ell_{o}}\Bigr],
\end{split}
\label{eq:BDwaveform}
\end{equation}
where recall that $\ell_{o}$ is related to the time at which the GWs are observed.
%$B_{k}(\bar{\alpha})$ is the Bessel-decomposed term of Eq. \ref{eq:BDsum} and we have included the complex conjugate terms as necessitated by Eq. \ref{eq:BD}.

%{\nl{update when you are done with the plots}} 

%In Fig.~\ref{fig:WF_time} we show an illustrative plot of the dispersion modulated time-domain waveform plotted against the GR waveform for three different GR-modified scenarios: massive gravitons, extra dimensions, and doubly special relativity. We reference Eq.~\eqref{eq:alpha_bar} in the plot titles to give clear values of $\bar{\alpha}$ which correspond to the changing values of luminosity distance which we set at $10$ Mpc, $100$ Mpc, and $1000$ Mpc for each different non-GR scenario. Fig.~\ref{fig:WF_time} serves as a useful visualization for the GW burst distortion in the time domain as a result of dispersion implementation where the severity and behavior of the distortion depend upon various parameters that we use to characterize $\bar{\alpha}$ including $\lambda_{\alpha}$ and $a$ which, again, define the strength and type of dispersion effect respectively.

\begin{figure*}[hbt!]
    \centering
    \includegraphics[width=\textwidth]{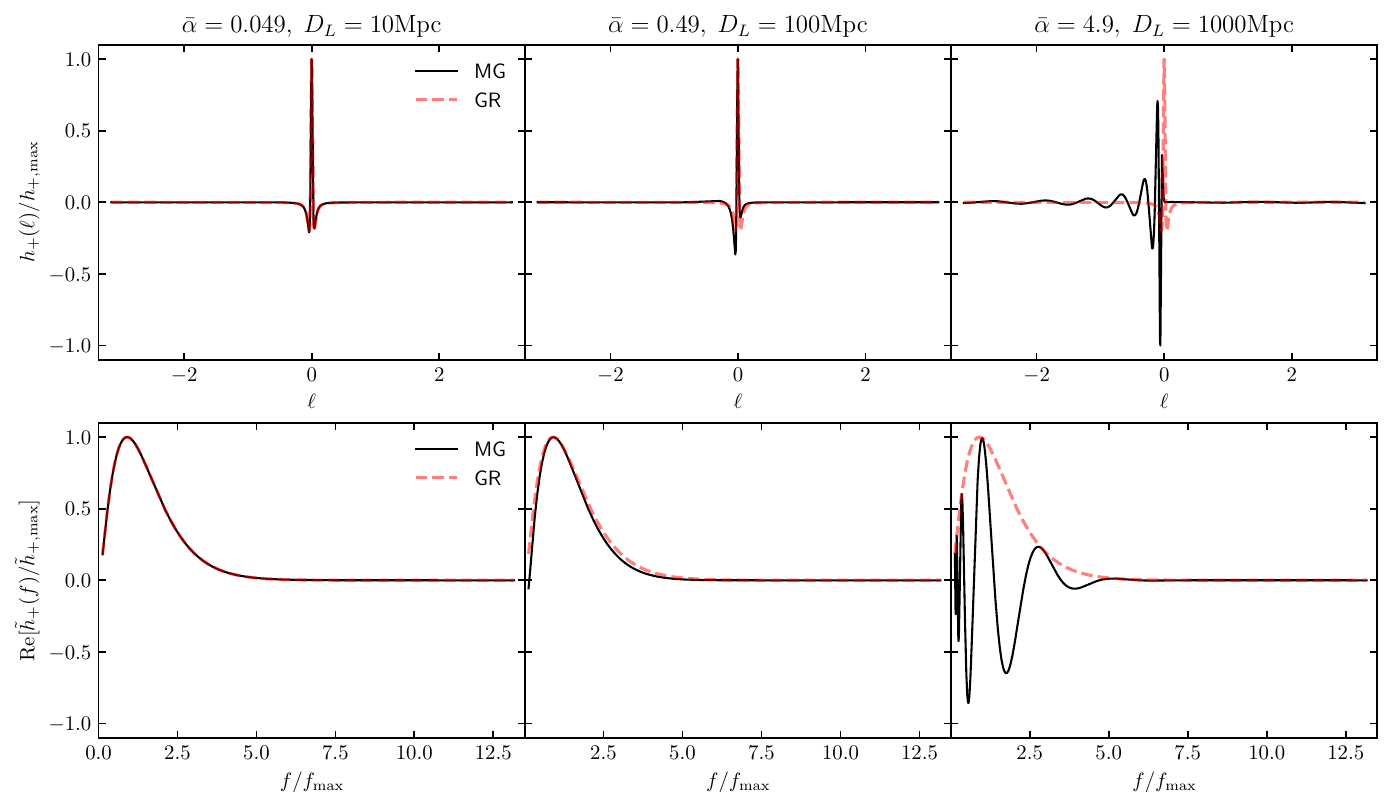}
    \caption{{\it Top:} Time-domain burst waveform from a binary with $m_{1} = 10 M_{\odot} = m_{2}$, $e=0.9$, and $n=9.4$ Hz subject to dispersion from massive gravitons $(a=0)$ at different distances from the source, specifically $D_{L} = [10, 100, 1000]$ Mpc (left, center, right). For this comparison, we have chosen $\lambda_{\rm MG}=1.2\times10^{-16} {\rm m}$ , which results in $\bar{\alpha}=[0.049,0.49,4.9]$ at the respective luminosity distances. The waveforms are plotted as functions of $\ell = n (t-t_{p})$ for simplicity, and are normalized by the maximum amplitude. The GR, non-dispersive, waveform (dashed line) is plotted for comparison. {\it Bottom:} The same waveforms but in the frequency domain, where $f_{\rm max} = nk_{\rm max}/2\pi$, and again, the waveforms are normalized by the maximum Fourier amplitude.}
    \label{fig:mg}
\end{figure*}

\begin{figure*}[hbt!]
    \centering
    \includegraphics[width=\textwidth]{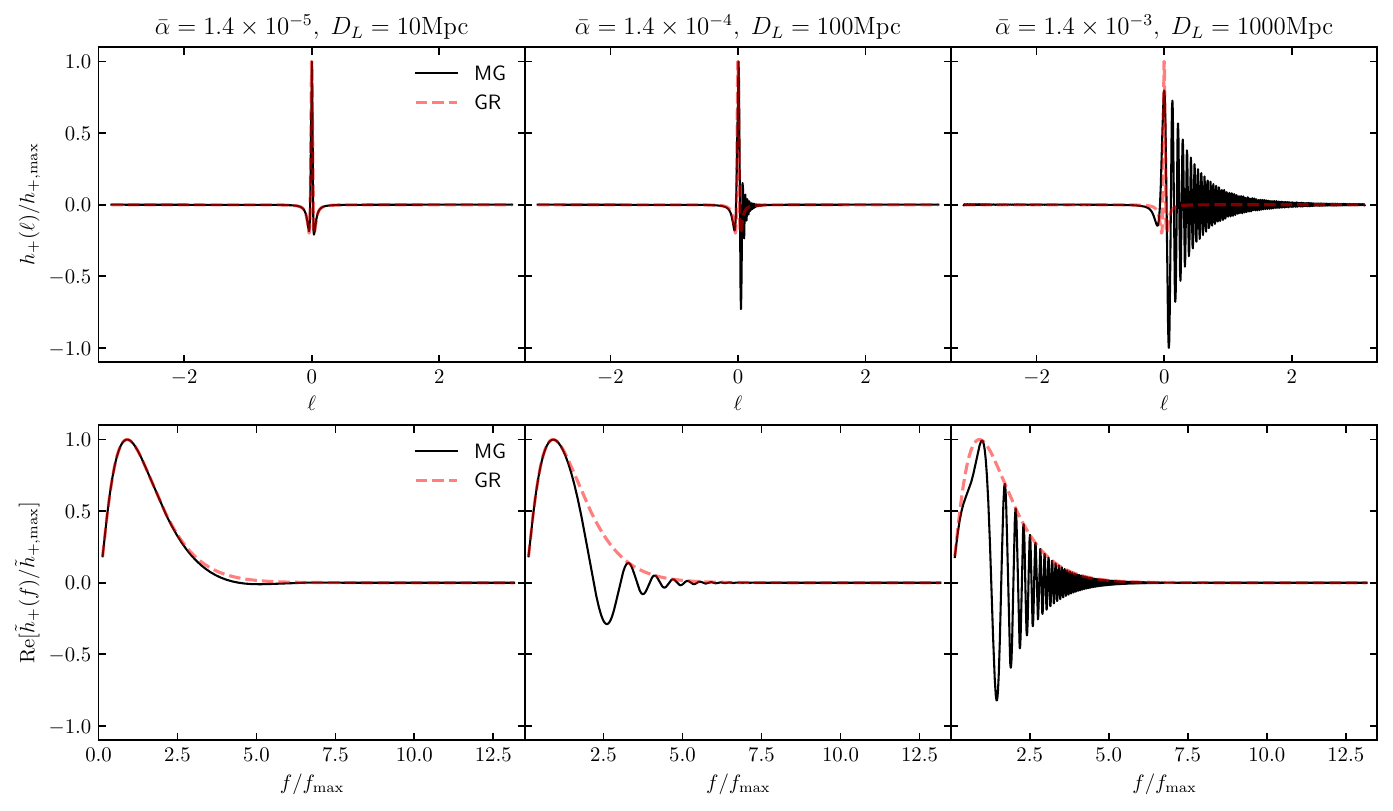}
    \caption{{\it Top:} Time-domain burst waveform from a binary with $m_{1} = 10 M_{\odot} = m_{2}$, $e=0.9$, and $n=9.4$ Hz subject to dispersion from extra dimensions $(a=4)$ at different distances from the source, specifically $D_{L} = [10, 100, 1000]$ Mpc (left, center, right). For this comparison, we have chosen $\lambda_{\rm edt}=1.0\times10^{-3}{\rm m}$, which results in $\bar{\alpha}=1.4\times[10^{-5},10^{-4},10^{-3}]$ at the respective luminosity distances. The waveforms are plotted as functions of $\ell = n (t-t_{p})$ for simplicity, and are normalized by the maximum amplitude. The GR, non-dispersive, waveform (dashed line) is plotted for comparison. {\it Bottom:} The same waveforms but in the frequency domain, where $f_{\rm max} = nk_{\rm max}/2\pi$, and again, the waveforms are normalized by the maximum Fourier amplitude.}
    \label{fig:ed}
\end{figure*}

We show an illustrative plot of the dispersion modulated time-domain waveform in the top panels of Figs.~\ref{fig:mg} and \ref{fig:ed}, for the massive graviton and extra dimension scenarios, respectively. We reference Eq.~\eqref{eq:alpha_bar} in the plot titles to give clear values of $\bar{\alpha}$ which correspond to the changing values of luminosity distance, specifically $[10,100,1000]$ Mpc.
%which we set at $10$ Mpc, $100$ Mpc, and $1000$ Mpc for each different non-GR scenario. 
We also plot the non-dispersive GR waveform (dashed lines) in each plot for comparison. Both of these figures reveal that dispersion causes an oscillatory modulation of the burst, before time of pericenter passage in the massive graviton case (Fig.~\ref{fig:mg}), and after for the extra dimensions case (Fig.~\ref{fig:ed}). 

Why does this behavior occur? The answer can be seen from Eq.~\eqref{eq:l_o}. For the massive graviton case, the disperive effect scales with negative powers of $k$. As a result, higher frequency components travel faster than lower frequency components, and would thus arrive sooner in a simulated detector reference frame. This causes a ``ringing'' in the early stages of the burst. For extra dimensions, and indeed for all cases with $a>2$, the opposite is true. High frequency harmonics travel slower than lower frequency harmonics, arriving later at a simulated detector, and causing the ringing to appear ``after'' the burst. It is worth noting that, at sufficiently high luminosity distance, the burst becomes completely smeared out to the point it no longer resembles a burst source, but instead resembles noise. However, in order to visualize this effect, one has to break the weak coupling assumption $\bar{\alpha} \ll 1$ in this case.

%these time-domain waveforms both under GR conditions for which $\bar{\alpha} = 0$ and for three different dispersive cases. For these waveforms we assume massive gravitons such that the change in $\bar{\alpha}$ corresponds to a change in luminosity distance $D_{L}$. Fig.~\ref{fig:WF_time} serves as a useful representation to understand how GWs in the time-domain are modulated by dispersion effects and deviate from the waveforms we model under GR.

% \begin{figure*}
%     \centering
%     \includegraphics[width=\textwidth]{hp_timedomain.pdf}
%     \caption{Comparisons of time-domain, normalized plus-polarization waveforms under GR and three different dispersive cases: massive gravitons (uppermost panel), extra dimensions (middle panel), and doubly special relativity (bottommost panel). The dashed, red line represents the GR waveform while the black solid lines show the different dispersive waveforms for different values of $\bar{\alpha}$ which assume different values as $D_{L}$ --- which we set at $10$ Mpc, $100$ Mpc, and $1000$ Mpc --- increases as denoted by the plot titles. The waveforms are computed using the following parameters: $e = 0.9$, $m = 20M_{\odot}$, $p = 20m$, $\eta = 0.25$}
%     \label{fig:WF_time}
% \end{figure*}

%------------------------------------------------
\subsection{Dispersion effects in frequency-domain eccentric burst waveforms}
\label{sec:efb}
%------------------------------------------------

The discussion of the previous section functions for both illustrative purposes, as well as to devise a general formalism of how to adjust for non-GR dispersion effects in eccentric PN waveforms. However, these are not the most useful for performing studies of projected constraints on non-GR effects, and we must move to the frequency domain to do so. This can be performed numerically, but to limit numerical error in our final results, we focus on implementing the dispersion effects into analytic frequency-domain eccentric waveforms. To do this, we break from the generality of the previous section, and restrict our attention to GW bursts from highly eccentric sources, which may be described by the effective fly-by (EFB) framework~\cite{Loutrel:2018ydv}. We will use a slightly modified EFB-F waveform which is derived in Appendix~\ref{app:efb-new} herein, which we will refer to as EFB-D waveforms.

The EFB-D waveforms in the absence of dispersion are obtained by double application of the stationary phase approximation (SPA), first to transform from the time-domain waveform in Eq.~\eqref{eq:time_polarizations}, and second to perform the resummation of the sum over $k$ in the frequency domain. To implement dispersion effects into this, we must Fourier transform Eq.~\eqref{eq:BDwaveform} which has the dispersion effect implemented in the time-domain polarizations, and determine whether either SPA is corrected by the introduction of such effects. First, the evolution of the binary under leading PN order radiation reaction is given by Eqs.~\eqref{eq:eoft}-\eqref{eq:loft} where $t$ is promoted to $t_{e}$, and subsequently, $\ell$ is promoted to $\ell_{e}$. One can rigorously show that these expressions do not change when mapping to $(t_{o}, \ell_{o})$, since $\partial t_{o}/\partial t_{e} = 1$ from Eq.~\eqref{eq:observed-time}. However, there is a simpler, and more physical, reason for this, specifically, Eqs.~\eqref{eq:eoft}-\eqref{eq:loft} are related to the generation of the waves from the binary, and thus, dispersion should not alter them.

The second application of the SPA, to perform the resummation over $k$, can be modified due to $B_{k}(\bar{\alpha})$ in Eq.~\eqref{eq:B_k}. If the dispersion effects are large, the phase of $B_{k}(\bar{\alpha})$ can dominate over the phase $\Psi_{\star}(k,f)$ in Eq.~\eqref{eq:Psi_star}. However, we are interested in parameterized tests of GR, for which the region of parameter space of interest is $\bar{\alpha} \ll 1$, since this is relevant dimensionless quantity for dispersion effects in eccentric binaries. Under this assumption, the dispersion effects should be small, and thus a slowly varying function compared to the orbital contribution to the phase from $\ell_{o}$. We can then use the generating function of Bessel functions of the first kind,
\begin{equation}
    e^{\frac{z}{2} (x - x^{-1})} = \sum_{m=-\infty}^{\infty} x^{m} J_{m}(z)\,,
\end{equation}
to recast the phase modulation of $B_{k}(\bar{\alpha})$ into an amplitude modulation, specifically
\begin{equation}
        B_{k}(\bar{\alpha}) = \sum_{m = -m_{\rm max}}^{m_{\rm max}} i^{m}\left(\frac{k}{k_{\rm max}}\right)^{m\left(1-\frac{a}{2}\right)} I_{m}\left(\bar{\alpha} k^{\frac{a}{2}}\right),
\label{eq:BDsum}
\end{equation}
where we have made use of the connection formula~\cite{1965hmfw.book.....A}
\begin{equation}
    I_{\nu}(x) = i^{-\nu} J_{\nu}(i x)\,,
\end{equation}
and we have introduced $m_{\rm max}$ to allow for the summation over $m$ to be truncated at a finite value. The observed time-domain waveform is still given by Eq.~\eqref{eq:BDwaveform}, but with $B_{k}(\bar{\alpha})$ now given by Eq.~\eqref{eq:BDsum}.

Mathematically, $m_{\rm max} \rightarrow \infty$ in order for Eq.~\eqref{eq:BDsum} to be an exact analytical representation of Eq.~\eqref{eq:B_k}. However, considering the numerical impossibility of evaluating such a sum over infinite bounds, we must find adequate limits on the summation to recover an accurate value while also maintaining computational viability. From the standpoint of tests of GR, $\bar{\alpha} \ll 1$, and since $I_{m}(\bar{\alpha}k^{a/2}) \sim \bar{\alpha}^{m}$ in this limit, one should choose $m_{\rm max} = 1$ when preforming tests of the null hypothesis. For more general cases where $\bar{\alpha}$ can take any arbitrary value, $m_{\rm max} > 1$ and must be chosen such that Eq.~\eqref{eq:BDsum} is a sufficiently accurate approximation of Eq.~\eqref{eq:B_k}. 

%To do this, we must consider how well the Bessel-decomposed waveform models the original waveform with the exact exponential term. The Bessel-decomposed waveforms include two infinite summations; one over $m$ and then one over $k$. As previously mentioned, infinite summation is not computationally possible, so we truncate both sums which, in the case of truncating the Bessel decomposition sum over $m$, causes the Bessel-decomposed waveform to deviate from the waveform with the exact exponential term. To find good bounds on $m$, we plot both the exact exponential term and the term derived from Bessel decomposition together up to some substantially large value of $k$ to see whether they give the same values for varying upper limits of the summation over $m$. We find that for large values of $k$ ($>200$), the Bessel-decomposed term exactly models the exact exponential for $m = 34$, so we perform our sum from $-34$ to $34$.

The frequency-domain waveform can now be obtained by computing
\begin{equation}
    \tilde{h}_{+,\times}^{(o)}(f) = \int dt_{o} h_{+,\times}(t_{o}) e^{2\pi i f t_{o}}\,,
\end{equation}
where $h_{+,\times}(t_{o})$ is given by Eq.~\eqref{eq:BDwaveform} with $B_{k}(\bar{\alpha})$ given in Eq.~\eqref{eq:BDsum}. The time integral over $t_{o}$ and resummation over $k$ are computed using the exact same methods as detailed in Appendix~\ref{app:efb-new}, thus we may write the final waveform directly,
\begin{equation}
    \label{eq:EFB}
    \tilde{h}^{(o)}_{+,\times}(f) = -\tilde{h}_{i} {\cal{A}}_{+,\times}(f) e^{2\pi i f t_{p}^{(o)}}\,,
\end{equation}
where
\begin{align}
    \tilde{h}_{i} &= \frac{2\pi}{D_{L}} \left(\frac{\cal{M}}{n_{i}}\right)^{1/3}\,,
    \\
    {\cal{A}}_{+,\times}(f) &= \left\{\lim_{k\rightarrow 2\pi f/n_{i}} \left[\Theta(k-1) H_{+,\times}^{(k)}(e_{i}) B_{k}(\bar{\alpha}) \right]^{\dagger}\right\}\,,
\end{align}
with ${\cal{M}} = m \eta^{3/5}$ the chirp mass of the binary, $n_{i} = 2\pi F_{{\rm orb},i}$ the mean frequency at pericenter, $e_{i}$ the orbital eccentricity at pericenter, $\Theta(x)$ the Heaviside step function, and $t_{p}^{(o)}$ a time offset that we will discuss in more detail in Sec.~\ref{sec:timing}.

In the bottom panels of Figs.~\ref{fig:mg} and \ref{fig:ed}, we show normalized plots of the dispersion-modulated $+$ polarization for the EFB frequency-domain waveforms at the same luminosity distances at the time-domain waveforms. Again, the non-dispersive GR waveform (dashed line) in the frequency domain is shown for comparison. This illustrative plot shows how the burst is also distorted in the frequncy domain by dispersion effects as the wave propagates and moves away from its source.

%By increasing the value of $\alpha$ in waveforms in both the frequency and time domains, Fig. \ref{fig:EFB_freq} shows that dispersion effects do distort the burst's propagation from its original state under GR (corresponding to smaller values of $\alpha$). Modifying $\alpha$ corresponds to changes in the luminosity distance $D_{L}$ under the assumption that the GW propagation is mediated by massive gravitons. Fig. \ref{fig:EFB_freq} shows how dispersion effects alter the GW propagation and distort the burst as luminosity distance increases and the burst moves away from its source. 

% \begin{figure*}
%     \centering
%     \includegraphics[width=\textwidth]{hp_freqdomain.pdf}
%     \caption{Comparisons of normalized plus component polarizations in the frequency domain under GR and using the EFB framework for three different dispersive cases: massive gravitons (uppermost panel), extra dimensions (middle panel), and doubly special relativity (bottommost panel). The dashed, red line represents the GR waveform while the black solid lines show the different dispersive waveforms for different values of $\bar{\alpha}$ which assume different values as $D_{L}$ --- which we set at $10$ Mpc, $100$ Mpc, and $1000$ Mpc --- increases as denoted by the plot titles. The waveforms are computed using the following parameters: $e = 0.9$, $m = 20M_{\odot}$, $p = 20m$, $\eta = 0.25$}
%     \label{fig:EFB_freq}
% \end{figure*}

%------------------------------------------------

%------------------------------------------------
\subsection{Modifications to time of arrival}
\label{sec:timing}
%------------------------------------------------

Under the EFB framework, the GW signal is modeled as one distinct burst over a finite time interval. To correctly model the orbital inspiral with multiple GW bursts, it is necessary to implement a timing model over which the EFB waveform evolves accounting for changes in eccentricity, semilatus rectum, and time of the GW signal arrival. Thus we must redefine the EFB waveform equation in Eq.~\eqref{eq:EFB} to account for such changes~\cite{Loutrel:2018ydv}, specifically
\begin{align}
    \tilde{h}_{+,\times}(f; M, \eta, p_{i}, e_{i}) &= -\tilde{h}_{i}  {\cal{A}}_{+,\times}(f; M, \eta, p_{i}, e_{i}) 
  %  \nn \\ 
  %  & \times 
  e^{2\pi i f t_{p,i}^{(o)}(M, \eta, p_{i}, e_{i}, t_{p,i-1}^{(o)})},
\label{eq:EFB_time_model}
\end{align}
where we have promoted the previously constant time shift $t_{p}^{(o)}$ to a function of the binary's parameters. Here, $i$ denotes the values of eccentricity $e$, semilatus rectum $p$, and time at which the signal is observed $t_{p}^{(o)}$ at the end of the $i^{\mathrm{th}}$-orbit which are related to the values of these quantities during the previous $i-1$ orbit by the following equations \cite{Arredondo:2021rdt, Loutrel:2019kky, Loutrel:2017fgu}:
\allowdisplaybreaks[4]
\begin{align}
    \label{eq:change_p}
    p_{i} &= p_{i-1} \Biggl[ 1 - \frac{128 \pi}{5} \eta \biggl( \frac{m}{p_{i-1}} \biggr)^{5/2} \biggl( 1 + \frac{7}{8} e_{i-1}^{2} \biggr) \Biggr]\,,
    \\
    \label{eq:change_e}
    e_{i} &= e_{i-1} \Biggl[ 1 - \frac{608 \pi}{15} \eta \biggl( \frac{m}{p_{i-1}} \biggr)^{5/2} \biggl( 1 + \frac{121}{304} e_{i-1}^{2} \biggr) \Biggr]\,,
    \\
    \label{eq:change_te}
    t_{p,i}^{(e)}  &= t_{p,i-1}^{(e)} + \frac{2 \pi}{m^{1/2}} \left[ \frac{p_{i-1}+\eta m\left(\frac{m}{p_{i-1}}\right)^{3/2}A(\epsilon_{i})}{\epsilon_{i-1}+\eta\left(\frac{m}{p_{i-1}}\right)^{5/2}B(\epsilon_{i})} \right]^{3/2}\,,
\end{align}
where $\epsilon \equiv 1-e^{2}$. Arredondo and Loutrel \cite{Arredondo:2021rdt} propose the following definitions for the functions $A$ and $B$,
\begin{align}
    \label{eq:A}
    A(\epsilon_{i}) &= a_{0} + a_{1}\epsilon_{i}\,,
    \\
    \label{eq:B}
    B(\epsilon_{i}) &= b_{0} + b_{1}\epsilon_{i} + b_{2}\epsilon_{i}^{2}\,.
\end{align}
The $a$ and $b$ coefficients have been calibrated against numerical evolution of the relative Newtonian order osculating equations to obtain~\cite{Arredondo:2021rdt}:
\begin{align}
    a_{0} &= a_{0,1} + a_{0,2}\left(\frac{\eta}{1/4}\right)^{a_{0,3}}\left(\frac{p}{10m}\right)^{a_{0,4}}\,,
    \\
    a_{1} &= -16.823395797589278\,,
    \\
    b_{0} &= 170\pi/3 =  178.0235837034216\,,
    \\
    b_{1} &= -139.3766232947201\,,
    \\
    b_{2} &= -1.088578314814299\,,
\end{align}
where
\begin{align}
    a_{0,1} &= 14.1774066465967\,,
    \\
    a_{0,2} &= -0.236903393660227\,,
    \\
    a_{0,3} &= 0.962439591179757\,,
    \\
    \label{eq:a04-coeff}
    a_{0,4} &= -2.415912280582671\,.
\end{align}
The above timing model given by Eqs.~\eqref{eq:change_p}-\eqref{eq:a04-coeff} models the evolution of the binary under relative Newtonian order radiation reaction (2.5PN quadrupole radiation). We implement these equations to adjust the time of pericenter $t_{p}$ and its time shift behavior to reflect the changing values of eccentricity and semilatus rectum due to the loss of orbital energy and angular momentum from GW emission.

When considering dispersion effects, the shift in pericenter passage time in Eq.~\eqref{eq:change_te} corresponds to a observer's clock at the location of the binary and does not account for any dispersion effects.  Thus we define the time variables as $t_{p}^{(e)}$ because this models the time as a measurement from the moment the burst is emitted. To introduce dispersion effects and obtain a timing model for $t_{p}^{(o)}$, we must again use the time shift in Eq.~\eqref{eq:observed-time} to relate $t_{p}^{(e)}$ in Eq.~\eqref{eq:change_te} to the observed time $t_{p}^{(o)}$ in Eq.~\eqref{eq:EFB_time_model}. To do this, we treat the bursts as discrete objects in time-frequency space with centroids $(t_{p,i}^{(o)}, f_{{\rm max},i})$. Thus, for general implementation of dispersive effects, the shift in the observed pericenter time of the burst is 
\begin{align}
    t_{p,i}^{(o)} &= t_{p,i-1}^{(o)} + \Delta t_{p}^{(e)} + \frac{D_{L}}{2 \lambda_{\alpha}^{2-a} f_{{\rm max},i-1}^{2-a}} g_{a}(f_{{\rm max},i}, f_{{\rm max},i-1})\,,
    \nn \\
    &= t_{p,i-1}^{(o)} + \Delta t_{p}^{(e)} + \frac{\bar{\alpha}_{i-1} k_{\rm max}^{a/2-1}}{2\pi F_{{\rm orb},i-1}} g_{a}(f_{{\rm max},i}, f_{{\rm max},i-1})
\label{eq:disp_timemodel}
\end{align}
where $\bar{\alpha}_{i}$ in the second equality is defined by evaluating Eq.~\eqref{eq:alpha_bar} on the $i$-th orbit, $\Delta t_{p}^{(e)} = t_{p,i}^{(e)} - t_{p,i-1}^{(e)}$ and is given in Eq.~\eqref{eq:change_te}, and we have defined
\begin{equation}
    \label{eq:g_a-def}
    g_{a}(f_{{\rm max},i}, f_{{\rm max},i-1}) = \left(\frac{f_{{\rm max},i-1}}{f_{{\rm max},i}}\right)^{2-a} - 1\,,
\end{equation}
with the subscript $a$ corresponding to the exponent of the non-GR dispersion effect.
%we now require the maximum harmonics of both the $i^{th}$ and $(i-1)^{\rm th}$ burst.

%The peak frequency $f_{\rm max}$ is a function of the eccentricity and semi-latus rectum through Eq.~\eqref{eq:kmax}, and thus evolves under time due to radiation reaction. The shift in this value between any two consecutive bursts will scale as a 2.5PN order correction. Thus, 
We may make a simplification to Eq.~\eqref{eq:disp_timemodel}, specifically to the dispersive factor $g_{a}$.
%\begin{equation}
%    g = \frac{1}{f_{\rm max,i}^{2-a}} - \frac{1}{f_{\rm max,i-1}^{2-a}},
%\label{eq:g}
%\end{equation}
%which we have defined as $g$ for simplicity in further references. 
This dispersive term $g_{a}$ can be expressed in terms of $k_{\rm max}$ in Eq.~\eqref{eq:kmax}, since $f_{\rm max} = k_{\rm max} F_{\rm orb}$. The $k_{\rm max}$ term is a function of $e_{i}$ which is in turn a function of $e_{i-1}$ through Eq.~\eqref{eq:change_e}. Likewise, $F_{{\rm orb},i}$ is a function of $(p_{i}, e_{i})$, which also maps to the previous values through the timing model. Thus, it is possible to perform a post-Newtonian (PN) expansion on $g_{a}$ by series expanding in Eq.~\eqref{eq:g_a-def} about $(m/p_{i-1}) \ll 1$. Then, $g_{a}$ becomes a function of $e_{i-1}$ and $p_{i-1}$ without $e_{i}$ and $p_{i}$ dependence. We choose to PN-expand the dispersive term because it simplifies the $g_{a}$ term as previously mentioned now solely depends on $i-1$ terms, which will make waveform parameter estimations simpler and calculations more efficient (Sec.~\ref{sec:params}). Performing the PN expansion to the first order yields
%{\nl{Ok, question here: When you did the PN expansion, did you also adjust for the fact that $F_{\rm orb}$ is different between the 2 orbits? If you didn't this will certainly impact the errors you get in Fig. 2.}} 
%
\begin{align}
    g_{a}(p_{i-1}, e_{i-1}) &= -\frac{2\pi}{15} \eta (a - 2)  \left(\frac{m}{p_{i-1}}\right)^{5/2} (1+e_{i-1})^{-1}
    \nn \\
    &\times 
 \left[-288 + 16(19\kappa-18)e_{i-1} - 252 e_{i-1}^{2}
  %  \right.
  %  \nn \\
  %  &\left.
    + (121\kappa - 252)e_{i-1}^{3}\right]
\label{eq:g_PN}
\end{align}
where $\kappa = 1.1954$.

%{\color{blue} I'm not sure we need the following information, but I included it for now!} 
%Fig.~\ref{fig:g_comp} shows a four-panel plot highlighting the behavior of $g_{\rm PN}$ in comparison to that of the exact $g$ term for three different binary systems and two dispersion effects. Table \ref{table:1} shows the exact parameters used to compute the values of $g$ and $g_{\rm PN}$. We see that for each binary type under both types of dispersion effect, the PN-expanded term closely follows the exact dispersive term with the relative error always staying below $25\%$ and averaging between $0.01\%$ and $0.02\%$ depending upon type of binary and dispersion effect. Considering the good similarity between the exact dispersive term and the PN-expanded term, we proceed using the PN expansion.

\begin{figure}
    \centering
    \includegraphics[width=0.6\columnwidth]{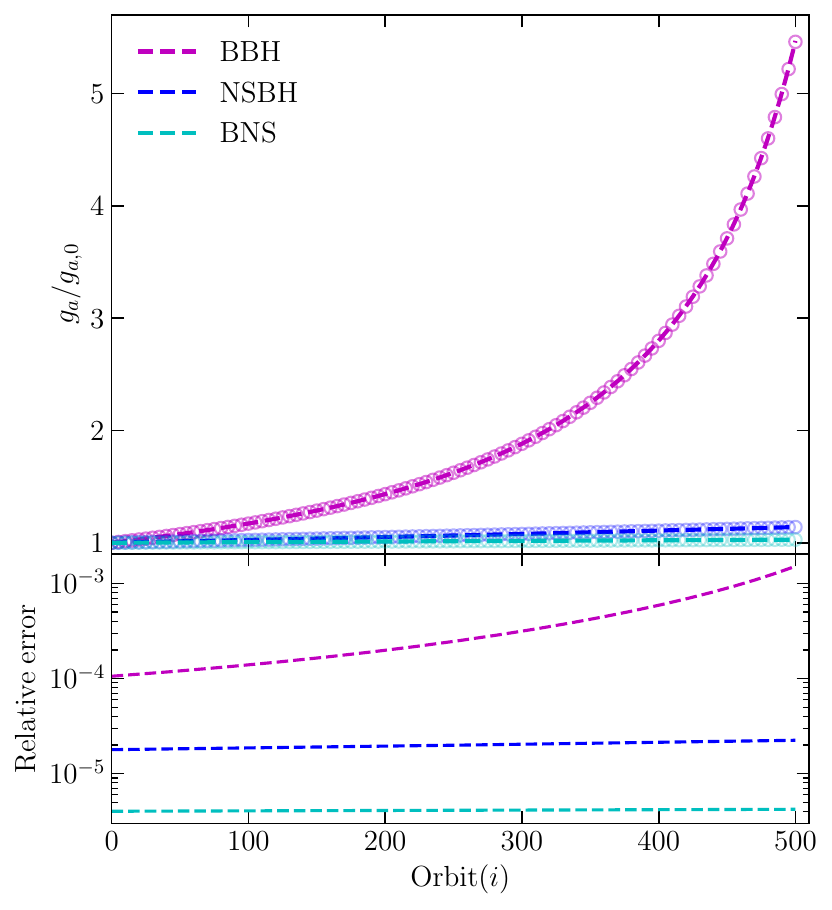}
    \caption{{\it Top:} Comparison between the exact expression for $g_{a}$ in Eq.~\eqref{eq:g_a-def} (circles) to its PN expansion in Eq.~\eqref{eq:g_PN} (dashed lines) for three different binaries (colors) with initial parameters given in Table~\ref{table:1}. Each binary is evolved using the burst mapping in Eqs.~\eqref{eq:change_p}-\eqref{eq:change_p} for five hundred orbits. Note that the combination $g_{a}/g_{a,0}$ is effectively independent of $a$ for the exact sequence, and actually independent in the case of the PN expansion in Eq.~\eqref{eq:g_PN}. {\it Bottom:} Relative error between the exact and PN expressions for $g_{a}$.}
    \label{fig:g_comp}
\end{figure}

Figure~\ref{fig:g_comp} shows a comparison between the PN expansion of $g_{a}$ in Eq.~\eqref{eq:g_PN} (dashed lines) and the exact expression in Eq.~\eqref{eq:g_a-def} (circles) for three different binary systems: a BBH (magenta), an NSBH (blue), and a BNS (red). Table~\ref{table:1} shows the orbital parameters for each of the binary systems. The exact expression and PN result show excellent agreement, with the relative error (bottom panel) being $\sim {\cal{O}}(10^{-5}-10^{-3})$ after five hundred orbits, depending on the system. Note that we plot the combination $g_{a}/g_{a,0}$, where $g_{a,0}$ is the initial value obtain from the orbital parameters in Table~\ref{table:1}. From Eq.~\eqref{eq:g_PN}, we see that this is independent of the value of $a$, and numerically we have checked that the same hold effectively true for the exact expression in Eq.~\eqref{eq:g_a-def}, despite its more complicated dependence on $a$. As a result, we do not show comparisons between different dispersion cases, i.e. the analysis in Fig.~\ref{fig:g_comp} is universal. %\dg{mention this universality again in the caption to fig 3?}

\begin{table}[t]
\centering
\begin{tabular}{c|c|c|c|c} 
 %\hline \hline
 System & $m1[M_{\odot}]$ & $m2[M_{\odot}]$ & $e(0)$ & $a(0)$ \\ [0.5ex] 
 \hline %\hline
 NS-NS & 1.4 & 1.4 & 0.9 & 1360 \\ %\hline 
 NS-BH & 1.4 & 10 & 0.9 & 535 \\ %\hline
 BH-BH & 10 & 10 & 0.9 & 368 \\ 
 %\hline \hline
\end{tabular}
\caption{Initial parameters used to evaluate Eq.~\eqref{eq:g_a-def} and Eq.~\eqref{eq:g_PN} for the binary systems considered in Fig.~\ref{fig:g_comp}.}
\label{table:1}
\end{table}

%shows a visual comparison of $g$ using the exact value dependent upon $e_{i}$ and $e_{i-1}$ with $g$ calculated from the PN expansion for different orbital parameters ranging from the initial conditions set at the $i = 0$ orbit up to the last stable orbit such that $p_{i} \ge 2m(3+e_{i})$. The PN expansion closely resembles the exact dispersive term with less than $20\%$ difference for each orbit. Because of the good similarity between the exact dispersive term and the PN expanded term, we proceed using the PN expansion.

% \begin{figure*}
%     \centering
%     \includegraphics[width=\textwidth]{g_comp.pdf}
%     \caption{Upper panels: Comparisons for massive gravitons and extra dimensions dispersion effects between exact EFB timing model dispersive term $g$ (solid dashed lines) and the PN-expanded terms $g_{\rm PN}$ (wider, lighter-colored lines) for three different binaries; binary neutron star (red), mixed neutron star/black hole (blue), binary black hole (magenta). The plots span from the $0^{\rm th}$ orbit with initial conditions in Table \ref{table:1} up to the last stable orbit such that $p_{i} \ge 2m(3+e_{i})$. Lower panel: Plot of the relative error in the PN-expanded terms.}
%     \label{fig:g_comp}
% \end{figure*}

%%%%%%%%%%%%%%%%%%%%%%%%%%%%%%%%%%%%%%%%%%%%%%%%%%%%%%%%%%%%%%%%%%%%%%%%%%%%%%%%%%%%%%%%%%%%%%%%%%%%%
\section{Parameter Estimation}
\label{sec:params}
%%%%%%%%%%%%%%%%%%%%%%%%%%%%%%%%%%%%%%%%%%%%%%%%%%%%%%%%%%%%%%%%%%%%%%%%%%%%%%%%%%%%%%%%%%%%%%%%%%%%

%-------------------------------------------------
\subsection{Fisher analysis for eccentric burst waveforms}
\label{sec:fisher}
%-------------------------------------------------

To study plausible constraints on modified dispersion effects with current and future GW detectors, we make use of the Fisher information matrix, an approximate method of obtaining parameter uncertainties (see Refs.~\cite{Vallisneri:2007ev,Porter:2015eha} for caveats of Fisher analyses). In the context of GWs, the Fisher method relies on taking derivative of, usually analytic, waveforms with respect to the model's underlying physical parameters to compute a covariance matrix, from which uncertainties can be extracted. Before considering the constraints on modified dispersion effects with eccentric burst sources, we provide the details of the general formalism of the Fisher analysis used for both single and repearted burst waveforms.

%----------------------------------
%\subsubsection{Single Burst Waveform Fisher Analysis}
%----------------------------------
The likelihood that, after subtracting a waveform model $h$, a detector data stream $s$ is purely given by noise is
\begin{equation}
    \ln {\cal{L}} = -\frac{1}{2} (s - h | s - h)\,,    
\end{equation}
where $(\; |\;)$ corresponds to the noise-weighted inner product %\dg{extrema of integration? I guess you're integrating over positive frequencies only, and so S is the one-sided PSD }
\begin{equation}
    (A | B) = 4 {\rm Re} \int_{f_{\rm low}}^{f_{\rm high}} \frac{df}{S_{n}(f)} A(f) B^{\dagger}(f)
\end{equation}
with $S_{n}(f)$ the one-sided power spectral density (PSD) for the GW detector in question, and $[f_{\rm low}, f_{\rm high}] = [10, 1024]$ Hz for the analysis carried out here. The waveform model will depend on a set of parameters $\theta^{a}$. Suppose that the detector data $s=h(\theta^{a}_{\rm true})$, where $\theta^{a}_{\rm true}$ are the ``true'' parameters of the source generating GWs, and the detector noise can be neglected. Under such an approximation, the likelihood around the $\theta^{a}_{\rm true}$ reduces to
\begin{equation}
    \ln {\cal{L}} = -\frac{1}{2} \Gamma_{ab} \Delta \theta^{a} \Delta \theta^{b} + {\cal{O}}[(\Delta \theta^{a})^{4})]\,,
\end{equation}
where $\Delta \theta^{a} = \theta^{a} - \theta^{a}_{\rm true}$, and
\begin{equation}
    \Gamma_{ab} = \left(\partial_{a} h | \partial_{b} h \right)
\end{equation}
is the Fisher information matrix with $\partial_{a} h = \partial h/\partial \theta^{a}$.  The variance $\sigma^{a}$ on a given parameter can be found by computing the inverse of the Fisher matrix, the covariance matrix $\Sigma_{ab} = [\Gamma^{-1}]_{ab}$, specifically
\begin{equation}
    \label{eq:variance}
    \sigma_{a} = \sqrt{\Sigma_{aa}}\,.
\end{equation}
Further, the correlation coefficient $c_{ab}$ between two parameters is given by the off-diagonal components of $\Sigma_{ab}$, i.e.
\begin{equation}
    c_{ab} = \frac{\Sigma_{ab}}{\sigma_{a} \sigma_{b}}\,.
\end{equation}
In the case of null tests of non-GR parameters, $\sigma_{a}$ determines the upper or lower bound on said parameters.

Before proceeding, there are a few extra details included in our analysis. First, the results of the Fisher analysis, specifically the variances $\sigma^{a}$ and correlation coefficients $c_{ab}$ can vary significantly depending on the sources sky location $(\theta_{S}, \phi_{S})$ and orientation with respect of the binary's angular momentum vector with respect to the line of sight $(\iota, \beta)$. To avoid this oscillatory dependence, we average over both these. The average of the sky localization reduces the inner produce to
\begin{align}
    \langle A | B \rangle &= \frac{1}{5} (A_{+} | B_{+}) + \frac{1}{5} (A_{\times} | B_{\times}) 
  %  \nn \\
  %  &
  = \frac{1}{5} \sum_{P=\{+,\times\}} \!(A_{P} | B_{P})\,,
\end{align}
while the average of the source orientation is more involved. From Eqs.~\eqref{eq:Spk}-\eqref{eq:Ckc}, we see that we may write
\begin{align}
    \label{eq:hplus-decomp}
    h_{+} &= (1 + \cos^{2}\iota)\left[ \cos(2\beta) h_{+}^{C} + \sin(2\beta) h_{+}^{S} \right] + \sin^{2}\iota \; h_{+}^{Q}\,,
    \\
    \label{eq:hcross-decomp}
    h_{\times} &= \cos\iota \left[\cos(2\beta) h_{\times}^{C} + \sin(2\beta) h_{\times}^{S}\right]\,,
\end{align}
where $h_{+,\times}^{C,S}$ and $h_{+}^{Q}$ do not depend on the orientation angles $(\iota,\beta)$. The average over $(\iota,\beta)$ results in
\begin{align}
    \label{eq:plus-avg}
    \langle A_{+} | B_{+}\rangle_{\iota,\beta} &= \frac{2}{15} \left[ 7 (A_{+}^{C} | B_{+}^{C}) + 4 (A_{+}^{Q} | B_{+}^{Q}) + 7 (A_{+}^{S} | B_{+}^{S})\right]\,,
    \\
    \label{eq:cross-avg}
    \langle A_{\times} | B_{\times} \rangle_{\iota,\beta} &= \frac{1}{6} \left[(A_{\times}^{C} | B_{\times}^{C}) + (A_{\times}^{S} | B_{\times}^{S}) \right]\,,
\end{align}
where we have assumed the polarizations $[A_{+,\times},B_{+,\times}]$ follow the decompositions in Eq.~\eqref{eq:hplus-decomp}-\eqref{eq:hcross-decomp}. Thus, the final expression for the averaged Fisher matrix we consider in this work is
\begin{equation}
    \Gamma_{ab}^{\rm avg} = \langle \partial_{a} h | \partial_{b} h\rangle = \frac{1}{5} \sum_{P=\{+,\times\}} \langle\partial_{a}h_{P} | \partial_{b} h_{P}\rangle_{\iota,\beta}
\end{equation}
where the average over $[\iota,\beta]$ is given by Eqs.~\eqref{eq:plus-avg}-\eqref{eq:cross-avg}.

%-----------------------------------------------
%\subsubsection{Multi-Burst Waveform Fisher Analysis}
%-----------------------------------------------
Another consideration comes from the nature of the emission itself, namely that highly eccentric binary systems emit GWs as repeated burst signals. The formalism above readily applies to single burst sources, but requires some modifications due to the fact that each subsequent burst is time-offset from the previous by an amount that depends on the parameters of the binary, which may be seen from the timing model in Eqs.~\eqref{eq:change_p}-\eqref{eq:g_PN}. Normally, this would not be an issue, since the total waveform would simply become a sum over the individual bursts, i.e
\begin{equation}
    \tilde{h}(f) = \sum_{j} \tilde{h}_{j}(f) e^{2\pi i f t_{p,j}(\theta^{a}_{j})}\,,
\end{equation}
with $\tilde{h}_{j}(f) = \tilde{h}(\theta_{j}^{a};f)$. The real challenge comes from the recursive nature of the burst timing model when confronted with the derivatives with respect to waveform parameters in the Fisher matrix. For single bursts, the waveform parameters are only those of the individual burst, so $\theta^{a}_{j}$. However, when considering a train of bursts, the entire sequence is only specified by the initial parameters $\theta^{a}_{0}$, and the Fisher analysis should be performed with respect to these, not the $\theta^{a}_{j}$ of the individual bursts. As a result, the waveform derivative in the case of multiple bursts becomes~\cite{Loutrel:2020jfx}
\begin{equation}
    \partial_{a_{0}} \tilde{h} = \sum_{j=0}^{N} {{\cal{J}}_{a_{0}}}^{a_{j}} \partial_{a_{j}} \tilde{h}_{j}
\end{equation}
where 
\begin{equation}
    {{\cal{J}}_{a_{0}}}^{a_{j}} = \prod_{k=1}^{j} \frac{\partial \theta^{a_{k}}_{k}}{\partial \theta^{a_{k-1}}_{k-1}}
\end{equation}
is the Jacobian associated with the mapping $\mu_{k}^{a}(\mu_{k-1}^{b})$ produced by the iterative timing model in Eq.~\eqref{eq:change_p}, \eqref{eq:change_te}, and \eqref{eq:disp_timemodel}, and $\partial_{a_{j}}h_{j} = \partial h_{j}/\partial \mu_{j}^{a_{j}}$. Note that for the first burst in the sequence, the Jacobian is simple the identity matrix.

The last consideration to note before presenting the results of the Fisher analysis is what parameters should be chosen as the waveform parameters $\theta^{a}$. It is well documented %at this point 
that eccentric burst waveforms contain significant parameter degeneracies, found first in studies of repeating burst sources~\cite{Loutrel:2020jfx} and later independently discovered in studies of unbound scattering encounters~\cite{Fontbute:2024amb}. The reason why is fairly straightforward: individual burst sources are not a full wave cycle, and thus, do not contain significant physical information. 

For the analysis carried out herein, we make use of the new Fourier domain EFB waveforms derived in Appendix~\ref{app:efb-new} and extended to include dispersion effects in Sec.~\ref{sec:efb}.
%, which are obtained by two applications of the stationary phase approximation (SPA). The waveforms, hereafter referred to as EFB-D, constitute leading PN order burst waveforms that can be combined with the timing model in Eqs.~\eqref{eq:change_p}-\eqref{eq:disp_timemodel} to obtain the waveform of a repeating burst sequence. 
Due to the fact that this waveform is only leading PN order, the degeneracies found in Ref.~\cite{Loutrel:2020jfx} are still present in this model, and hence the parameters for the Fisher analysis should be $\theta^{a} = [{\cal{M}}, {\cal{P}}_{0}, e_{0}, t_{p,0}, D_{L}, \bar{\alpha}_{0}]$, where ${\cal{P}} = (p^{3}/M)^{1/2}$ is the ``radius of curvature'' of the orbit, and the underscore $0$ indicates these are the initial parameters of the burst sequence, i.e. those of the first burst. However, when performing the Fisher analysis on the first burst, an immediate degeneracy appears. Specifically, $t_{p,0}$ is independent of any other parameter, while the waveform will only depend ${\cal{M}}$ and $D_{L}$ through the amplitude constant $\tilde{h}_{0}$. Thus, for single burst waveforms, these two parameters cannot be measured independently due to this degeneracy, but $\tilde{h}_{0}$ can be measured. As a result, the parameters that we actually use for the Fisher analysis are $\theta^{a} = [{\cal{M}}, {\cal{P}}_{0}, e_{0}, t_{p,0}, \tilde{h}_{0}, \bar{\alpha}_{0}]$, and priors in these parameters are chosen to be flat so that the posterior probability distribution is proportional to the likelihood. The EFB-D waveforms can be mapped into this parameterization by suitable transformation of variables on the timing model in Eqs.~\eqref{eq:change_p}-\eqref{eq:disp_timemodel}, and using the relationship $n_{i} = (1-e_{i}^{2})^{3/2}/{\cal{P}}_{i}$ in the waveform amplitudes.

\subsection{Constraints of modified dispersion}
\label{sec:constraints}
%-------------------------------------------------

To obtain projected bounds one might obtain on modified dispersion effects, we consider two representative binary sources for LIGO, specifically $m_{1} = m_{2} = 10M_{\odot}$ and $m_{1} = m_{2} = 30M_{\odot}$. The model used herein only depends on the chirp mass, so we do not consider binaries of different mass ratios. For the two binaries in question, we vary the initial eccentricity in the range $[0.8,0.99]$ and fix the peak frequency of the initial burst of the sequence to be $f_{\rm max} =40$ Hz. This is to ensure that the individual bursts would be approximately detectable via a burst search algorithm (see Fig. 1 of Ref.~\cite{Loutrel:2020jfx}). For this value of the peak frequency, the signal-to-noise ratio (SNR) of the initial burst for the lighter binary is $\rho\sim 7$, while for the heavier binary $\rho \sim 45$. The luminosity distance for all computations herein is fixed at $D_{L} = 100$ Mpc. Since $D_{L}$ does not explicitly enter the Fisher analysis, varying this parameter primarily affects the SNR $\rho$, which changes the uncertainties in a Fisher analysis due to the known fact that $\sigma^{a} \sim \rho^{-1}$. For each initial eccentricity value, we evolve each binary using the timing model up to a final eccentricity of $e_{f}=0.70$, ensuring that the GW signal is still burst-like and the approximations used herein remain valid. All inner products are numerically integrated from $f_{\rm low} = 10$ Hz to $f_{\rm high} = 1024$ Hz, and we use the theoretical design sensitivity curve for LIGO available at Ref.~\cite{Barsotti-LIGO} to compute $S_{n}(f)$. 

\begin{figure*}[t]
    \centering
    \includegraphics[width=0.495\columnwidth]{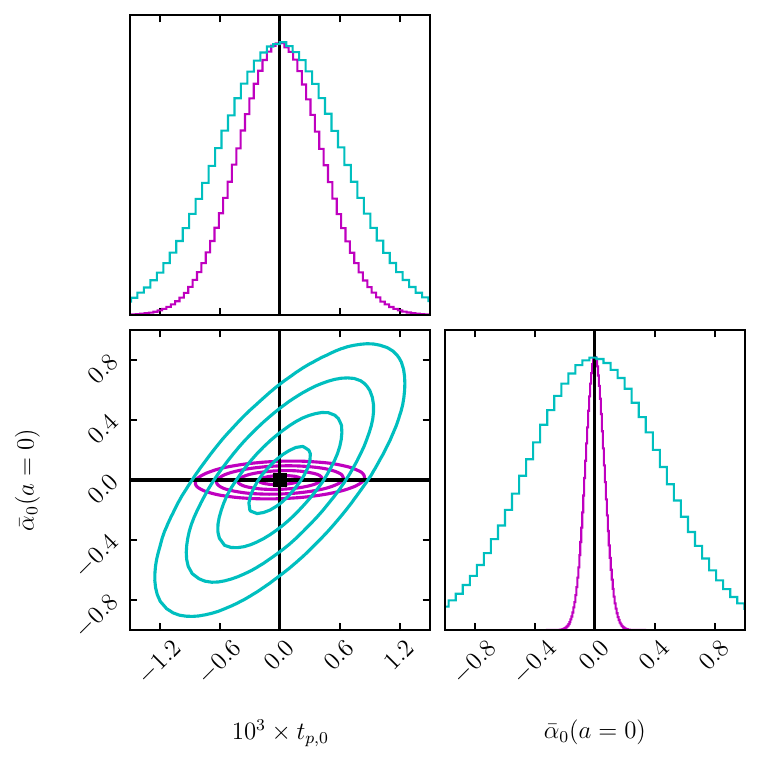}
    \includegraphics[width=0.495\columnwidth]{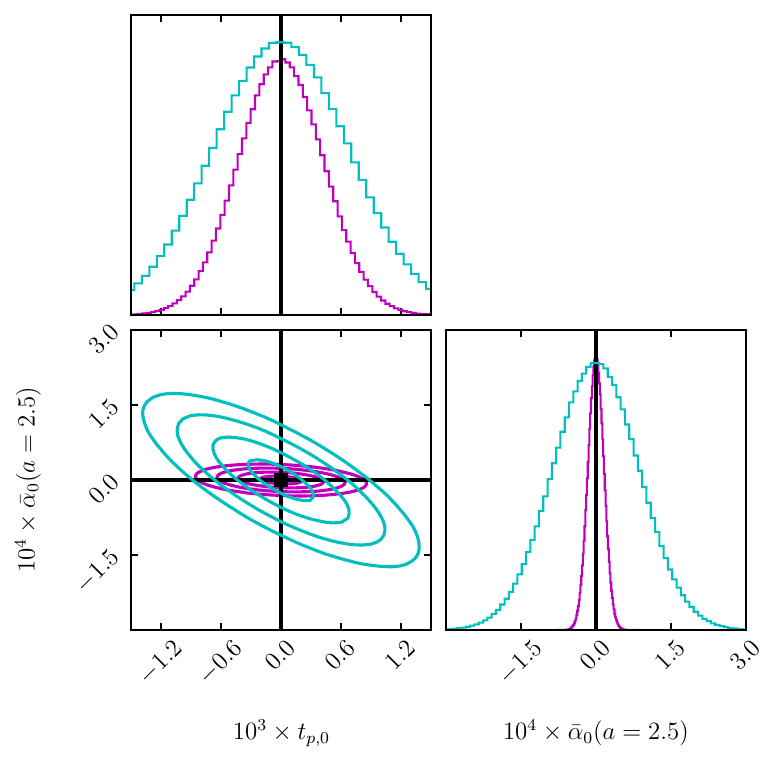}
    \includegraphics[width=0.495\columnwidth]{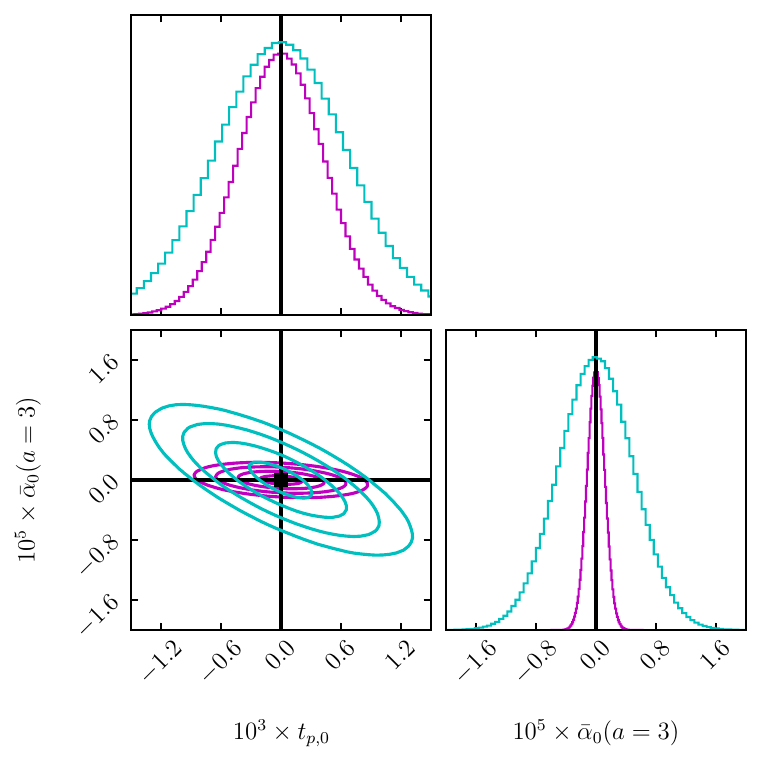}
    \includegraphics[width=0.495\columnwidth]{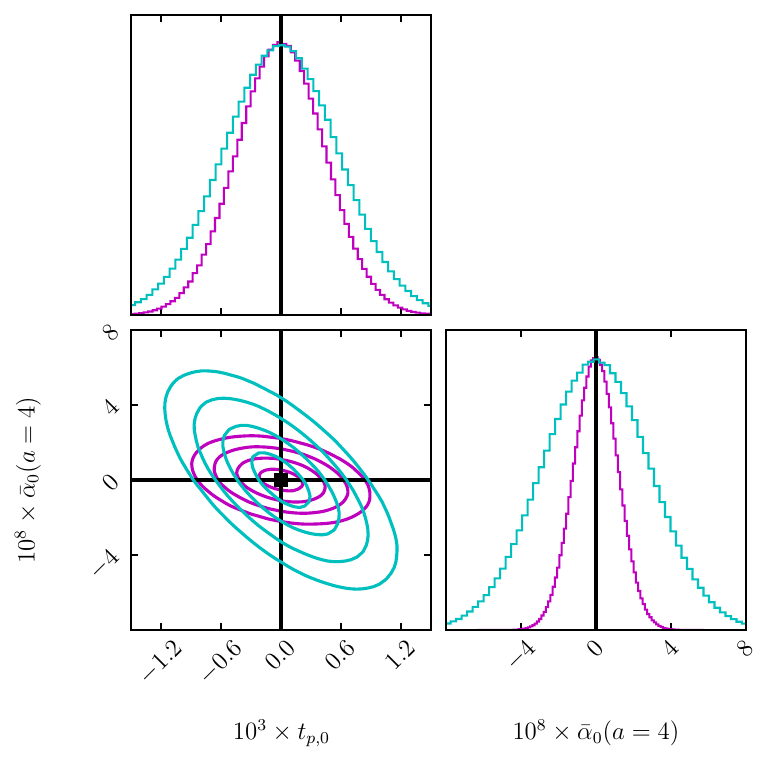}
    \caption{Marginalized posterior distributions for the binary with $m_{1} = m_{2} = 10M_{\odot}$ and $e_{0}=0.99$, for one burst (cyan) and a sequence of forty-eight bursts ending at $e_{f} = 0.70$ (magenta). The posteriors are shown for different values of the dispersion exponent parameter, thus changing the theoretical mechanism: $a=0$ (top left), $a=2.5$ (top right), $a=3$ (bottom left), and $a=4$ (bottom right). For the single burst event, the dispersion parameter $\bar{\alpha}_{0}$ is highly correlated with the time of pericenter of the burst $t_{p,0}$. Further, because of the degeneracies in the waveform, the luminosity distance and chirp mass cannot be measured independently, and thus any bounds on $\bar{\alpha}_{0}$ cannot be mapped to bounds on specific theoretical mechanisms. For the multi-burst sequence, the correlation between $\bar{\alpha}_{0}$ and $t_{p,0}$ is weakened, and the measurement uncertainty on the former improves. Further, the degeneracy between ${\cal{M}}$ and $D_{L}$ broken, and the bounds on $\bar{\alpha}_{0}$ can then be mapped to bounds on the theory specific mechanisms in Table~\ref{tab:summary}. Contours specify the $(0.5, 1, 1.5, 2)-$sigma credible levels of each distribution.}
    \label{fig:corner}
\end{figure*}

For the initial burst of the sequence, the Fisher analysis, more specifically the inversion of the Fisher matrix, is ill-posed, due to the initial burst's independence of the chirp mass. However, the Fisher matrix can be inverted by eliminating the corresponding row and column from the matrix. Doing so reveals reasonable uncertainties on $t_{p,0}$ and $\bar{\alpha}_{0}$ while the uncertainties for the other parameters remain large, which is consistent with previous results~\cite{Loutrel:2020jfx}. In Fig.~\ref{fig:corner}, we plot marginalized posteriors generated from the covariance matrix for the lighter binary with $e_{0}=0.99$, and for varying values of the exponent parameter $a$. For the posteriors of the initial burst (cyan contours), $\bar{\alpha}_{0}$ shows significant correlation with the time offset $t_{p,0}$. There is no strong correlation with any of the other waveform parameters. For comparison, the posteriors of the full burst sequence (magenta contours) is also displayed. For this initial eccentricity value, there are forty-eight bursts in the full sequence, after which the covariance between $\bar{\alpha}_{0}$ and $t_{p,0}$ is effectively broken, and the uncertainty on the former improves significantly. The uncertainty on the latter is largely set by the maximum frequency sampled, and thus, only improves marginally throughout the sequence.
\begin{figure*}[t]
    \centering
    \includegraphics[width=\textwidth]{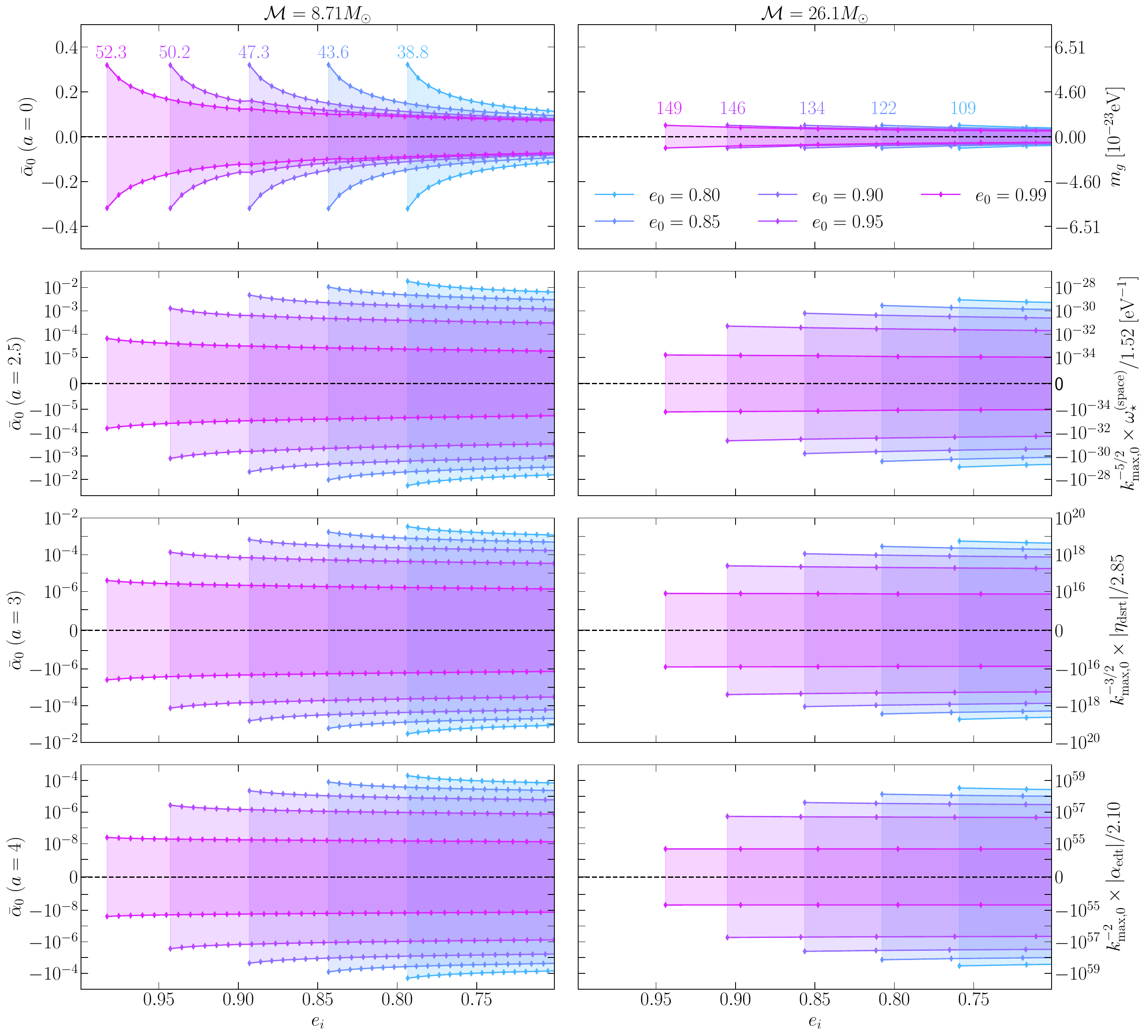}
    \caption{Uncertainty on the parameter $\bar{\alpha}_{0}$ as a function of eccentricity for a sequence of bursts, starting at $e_{0} = [0.99, 0.95, 0.90, 0.85, 0.80]$ and ending at $e_{f} = 0.70$. Left and right panels display two different binaries, $m_{1}=10M_{\odot}=m_{2}$ and $m_{1} = 30M_{\odot}=m_{2}$, respectively. From top to bottom, the panels corresponds to different dispersion scenarios with $a=[0,2.5,3,4]$, respectively. The right axes show the projection of the uncertainty on $\bar{\alpha}_{0}$ into theory specific bounds for a subset of the mechanisms in Table~\ref{tab:summary}, specifically massive gravitons ($a=0$), timelike multifractional spacetime ($a=2.5$), doubly special relativity ($a=3$), and extra dimensions ($a=4$). The theory specific bounds are multiplied by a factor of $k_{{\rm max},0} = k_{\rm max}(e_{0})$ to ensure that the mapping from the values of $\bar{\alpha}_{0}$ on the left axis are eccentricity independent. For more details, see Sec.~\ref{sec:constraints}. Colored numbers next to each curve in the top panels display the total SNR of the burst sequence. Note that the SNR does not change when considering different dispersion effects.}
    \label{fig:fisher}
\end{figure*}
\begin{figure*}[t]
    \centering
    \includegraphics[width=\textwidth]{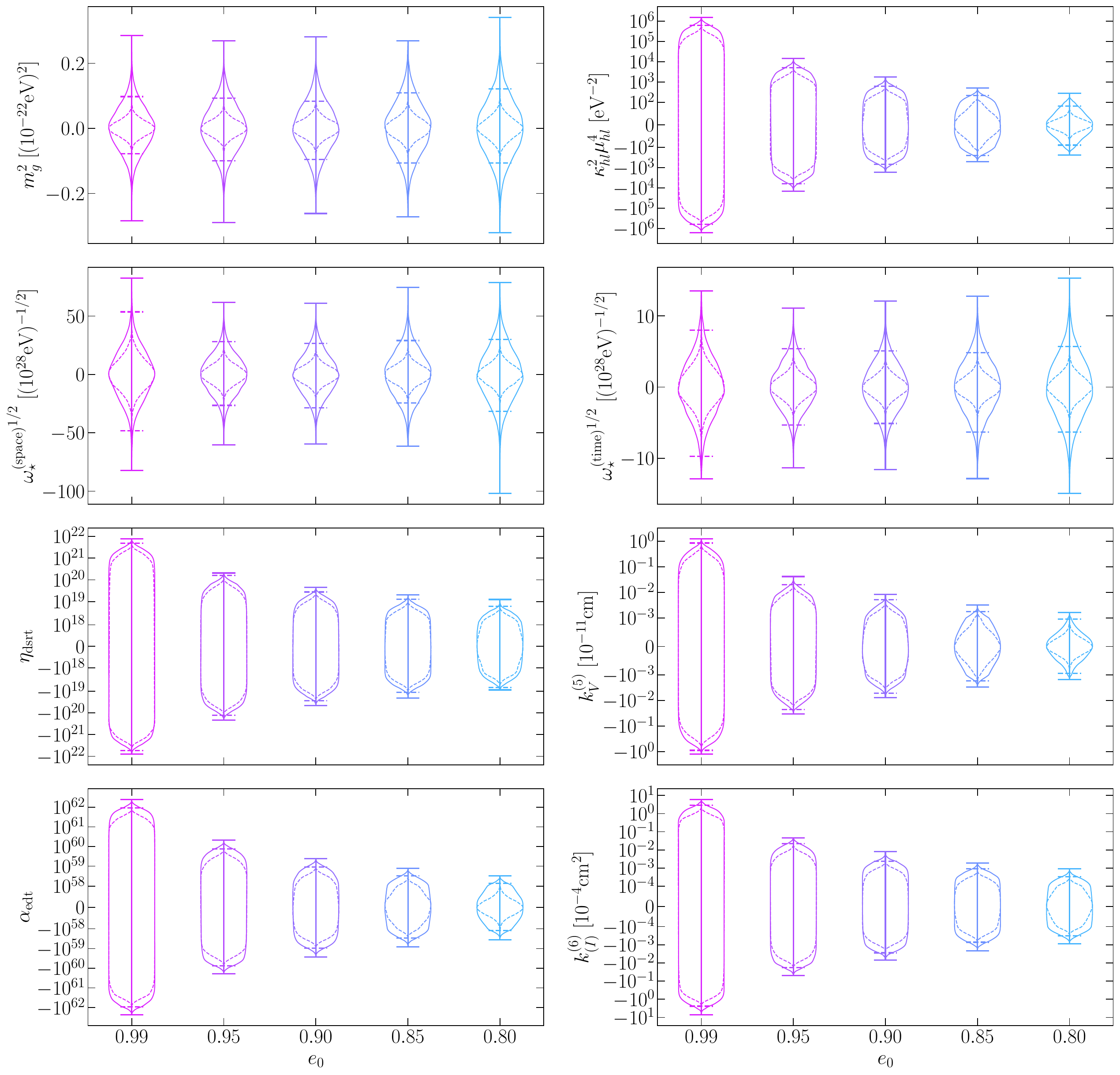}
    \caption{Violin plots on theory parameters found from taking the uncertainties on $\bar{\alpha}_{0}$ obtained from Fig.~\ref{fig:fisher} and using the mapping in Eq.~\eqref{eq:disp-bound}, along with the appropriate expression for $\alpha$. The latter can be found in Sec.~\ref{sec:theory} as well as Table~\ref{tab:summary}. Solid distributions correspond to results for the binary with chirp mass ${\cal{M}}=8.71 M_{\odot}$ ($m_{1}=m_{2}=10M_{\odot})$, while the dashed distributions are for a binary with ${\cal{M}}=26.1{\cal{M}}_{\odot}$ ($m_{1}=m_{2}=30M_{\odot}$). Each plot corresponds to a different theoretical mechanism, specifically massive gravitons (top row, left), Horava-Lifschitz gravity (top row, right), multifractional spacetimes (upper middle row, left and right), doubly special relativity (bottom middle row, left), gravitational SME for $d=5$ (bottom middle row, right) and $d=6$ (bottom row, right), and lastly extra dimensions (bottom row, left).}
 %\dg{These are Fishers, so the posterior are Gaussian by definition. The violin plots in this figure are not Gaussians though, they are very flat at the top. With a log-scale on the y axis, those violin plots should look like sideways parabolas (log of Gaussian = quadratic polynomial). Do you see what I mean? Fig 4 has nice Gaussians, perhaps this just is a plotting artifact? }}
 
%\dg{You could add labels in the various panels indicating which modification one is looking at. For instance "massive gravitons", etc. This  applies to all figures in the paper (... and feel free to ignore if you don't want to re-do them, I just thought it would be useful when putting these plots in one's slides for a talk)}
    \label{fig:violins}
\end{figure*}

For the binaries studied here, Fig.~\ref{fig:fisher} shows the uncertainty on $\bar{\alpha}_{0}$ (colored diamonds) plotted against $e_{i}$, showcasing how the uncertainty changes as the burst sequence proceeds for different values of the dispersion exponent parameter $a$. Shaded regions correspond to points contained within the 1-sigma contours as determined by the uncertainty computed through Eq.~\eqref{eq:variance}. The sequence starts from the second burst due to the ill-posedness of the matrix inversion on the first burst. In general, the uncertainty on $\bar{\alpha}_{0}$ decreases as the sequence proceeds. Higher chirp mass and higher initial eccentricity, save for the case $a=0$, lead to better bounds on $\bar{\alpha}_{0}$.

The uncertainty on $\bar{\alpha}_{0}$, specifically $\sigma_{\bar{\alpha}_{0}}$, can be mapped into theory specific bounds on the coupling parameters using Eq.~\eqref{eq:alpha_bar} for the initial burst, i.e.
\begin{equation}
    \label{eq:disp-bound}
    k_{{\rm max},0}^{-a/2} \; \alpha \le \frac{(2\pi)^{1-a}}{D_{L}} \sigma_{\bar{\alpha}_{0}} L_{P}^{4-2a} f_{{\rm max},0}^{1-a}
\end{equation}
where $[k_{{\rm max},0}, f_{{\rm max},0}, D_{L}]$ can be inferred from the recovered parameters $[e_{0}, {\cal{P}}_{0}, \tilde{h}_{0}, {\cal{M}}]$. As a reminder, all of the binaries considered here have $f_{{\rm max},0} = 40$ Hz and $D_{L}=100$ Mpc, hence the only parameter that vary in the above equation are $k_{{\rm max},0}$ and $\sigma_{\bar{\alpha}_{0}}$. Inserting the uncertainty on $\bar{\alpha}_{0}$ from the results of the Fisher analysis into Eq.~\eqref{eq:disp-bound} gives the bound on theory specific parameters. For Fig.~\ref{fig:fisher}, the right axis displays this specific mapping for a subset of the theoretic mechanisms in Table~\ref{tab:summary}. Note that the combination $k_{{\rm max},0}^{-a/2} \alpha$, or more specifically the right-hand-side of Eq.~\eqref{eq:disp-bound}, is independent of the initial eccentricity.

Naively, one might expect from Fig.~\ref{fig:fisher} that the strongest bounds come from the binaries with the highest initial eccentricity $e_{0}$. However, this is not necessarily true since $k_{{\rm max},0} \sim (1-e_{0}^{2})^{-3/2}$, and can thus vary by several orders of magnitude as $e_{0}\rightarrow 1$. To showcase the theory specific bounds obtained from the complete burst sequences, we simulate violin plots in Fig.~\ref{fig:violins} by generating normal distributions using the uncertainty on $\bar{\alpha}_{0}$ with the correct theory specific conversion factors. Solid line correspond to result from the lighter binary, while dashed lines correspond to the heavier binary. For three of the cases, namely massive gravitons (top left), and space-like (upper middle left) and time-like (upper middle right) multifractional spacetimes, the bounds on the theory parameters do not change significantly with the initial eccentricity $e_{0}$. The remaining cases, specifically doubly special relativity (lower middle left), extra dimensions (bottom left), Ho\v{r}ava-Lifschitz gravity (top right), and gravitational standard model extensions (lower middle and bottom right), show significant variation over several orders of magnitude in the theory parameters. In all of these cases, $e_{0}=0.80$ gives the best constraint of the systems studied here. Further, the heavier binary generally gives better constraints regardless of the theoretical mechanism considered, but this is due to the increase in SNR compared to the lighter binary.

As a final piece of our analysis, we repeat the Fisher computation for an ensemble of $10^{3}$ binaries at $D_{L} = 100$ Mpc. To obtain the ensemble, we randomly select the chirp masses and peak frequencies of the initial burst uniformly in the ranges ${\cal{M}} \in [8,80]M_{\odot}$ and $f_{\rm max} \in [10,30]$ Hz. We start all of the binaries on initially parabolic orbits $(e_{0}=1)$, thus constituting a set of dynamical capture binaries that become bound due to GW emission during the first pericenter passage. These three values in hand fix ${\cal{P}}$ and all subsequent waveform parameters. We then allow all of the binaries to evolve up to eccentricity $e=0.7$, and stop the analysis there to ensure that we are still under the burst assumptions.

\begin{figure}
    \centering
    \includegraphics[width=0.6\columnwidth]{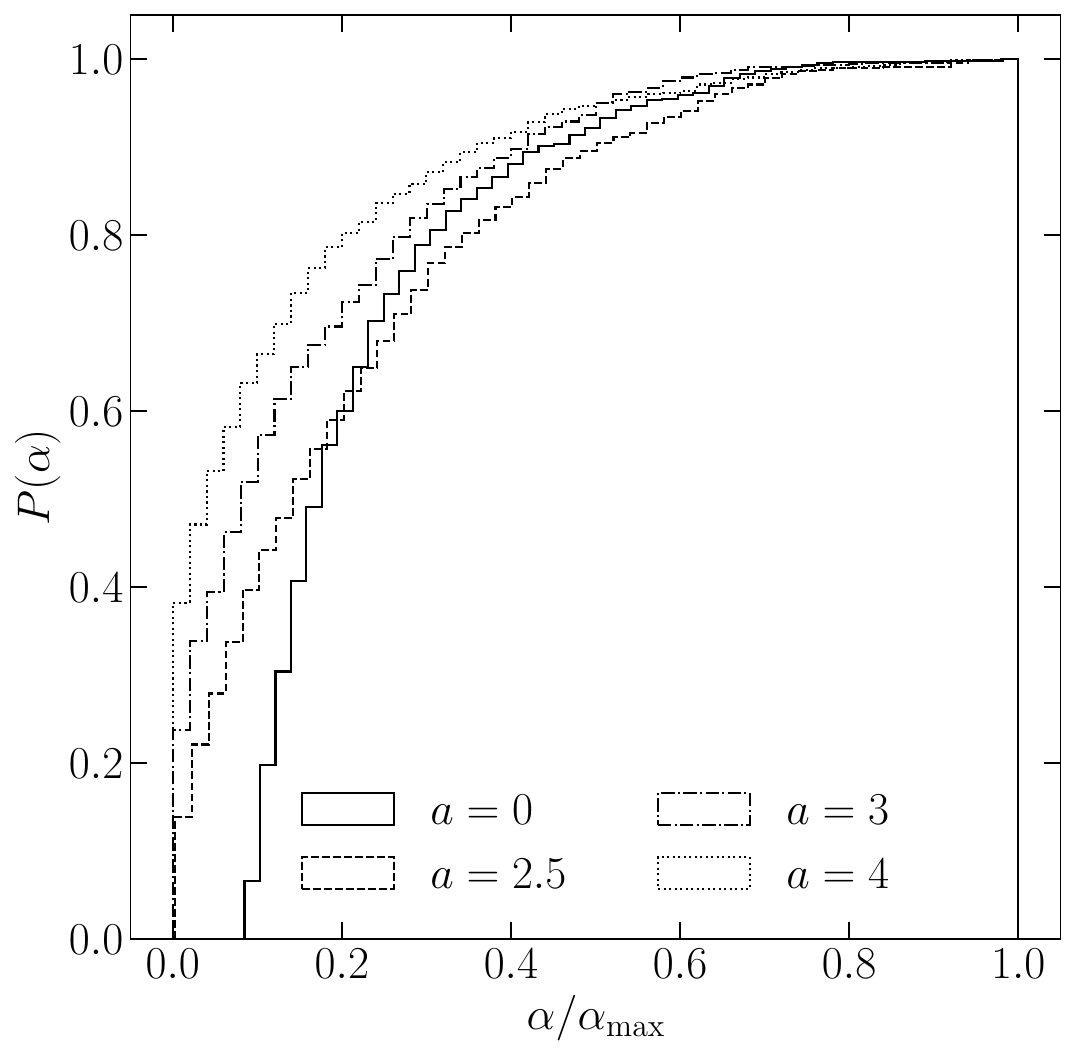}
    \caption{Cumulative probability distribution of the bound on $\alpha$ for an ensemble of one thousand GW capture binaries for different exponent parameters $a$. The ensemble is chosen such that ${\cal{M}}\in[8,80]{\cal{M}}_{\odot}$ and $f_{\rm max} \in [10,30]$ Hz, with $e_{0}=1$ and $D_{L} = 100$ Mpc. The distributions are plotted versus $\alpha/\alpha_{\rm max}$, where $\alpha_{\rm max}$ corresponds to the largest bound obtained from the ensemble. This is simply to normalize all bounds for different values of $a$ so they appear on the same plot.}
    \label{fig:ensemble}
\end{figure}

The analysis of the ensemble gives us bounds on the theory agnostic parameter $\bar{\alpha}$ for any given $a$, which can then be mapped to bounds on the theory specific parameter by Eq.~\eqref{eq:disp-bound}. Thus, we obtain a distribution $p(\alpha)$ for the bounds on $\alpha$. We plot the cumulative distribution function $P(\alpha)$ for this distribution in Fig.~\ref{fig:ensemble} as a function of $\alpha/\alpha_{\rm max}$, where $\alpha_{\rm max}$ is the largest (weakest) bound obtained from the ensemble of binaries. %We plot it this way due to the fact that the 
Note that theory-specific bounds can vary by many orders of magnitude depending on the specific scenario and the value of $a$. Figure~\ref{fig:ensemble} shows that eighty percent of the binaries in our ensemble provide bounds $\alpha_{<80\%}\lesssim [0.25,0.32,0.20,0.15]\alpha_{\rm max}$. As an example, this implies that eighty percent of the binaries in our sample will produce bounds on the graviton mass $m_{g} \lesssim 4.8\times 10^{-23} {\rm eV}$. The equivalent values for the remainder of the theory parameters are given in the last column of Table~\ref{tab:summary}.

Table~\ref{tab:summary} also contains the constraints obtained from analysis of GWTC-3~\cite{LIGOScientific:2021sio} mapped into the theory specific parameters for each scenario. An important thing to note regarding the constraints from eccentric burst sources is that we fixed the luminosity distance, which is useful to interpret our results but cannot be directly compared with detection catalogs. Nonetheless, a more reasonably comparison can be made by rescaling our results by the average SNR of the sources. For our ensemble, the average SNR is approximately 140, while for GWTC-3, the average is approximately 11. Since uncertainties in a Fisher analysis scale inversely with SNR, the constraints in the last column would get weaker by an order of magnitude when rescaling to the average SNR of the GWTC-3 events. This indicates that dispersion scenarios with smaller $a$ value, massive gravitons and multifractional spacetimes for example, would perform worse compared to equivalent constraints from quasi-circular binaries. On the other hand, the larger $a$ is, i.e. extra dimensions and the gravitational SME, the better our results are compared to equivalent quasi-circular bounds. 

%

%%%%%%%%%%%%%%%%%%%%%%%%%%%%%%%%%%%%%%%%%%%%%%%%%%%%%%%%%%%%%%%%%%%%%%%%%%%%%%%%%%%%%%%%%%%%%%%%%%%%%
\section{Discussion}
\label{sec:disc}
Eccentricity has a non-trivial impact on our ability to probe modified GW dispersion effects. Figure~\ref{fig:violins} and Table~\ref{tab:summary} shows how this depends heavily on which type of effect one is considering. For effects such as massive gravitons and multifractional spacetimes, the projected bounds obtained here are not better than those obtained from the population-level analysis of GWTC-3~\cite{LIGOScientific:2021sio}. This naively seems to indicate that current detectors have already hit the limit of how well these dispersion effects can be constrained with single events. This further implies that improvements over current bounds can only be obtained by building more sensitive dectors or combining inference from multiple events, such as those found with GWTC-3~\cite{LIGOScientific:2021sio}.

On the other hand, for effects with dispersve exponent parameters $a\ge3$ (Ho\v{r}ava-Lifschitz gravity, extra dimensions, gravitational SME, doubly special relativity, etc.), improved bounds can be obtained from eccentric burst sources under optimal conditions. These improvements can be an order of magnitude as in the case of doubly special relativity, and up to three orders of magnitude for the $d=6$ gravitational SME. The reason for this is that, for $a>2$, the time of arrival of different harmonics within the burst, as well as the difference in times of arrival of the bursts themselves, grows like $\sim f^{a-2}$, where here $f$ is the relevant frequency value. Hence, the larger the value of $a$, the more small changes to the dispersion coupling parameter will have a large impact on the propagation of GWs.

There are a few things to note about the results in Fig.~\ref{fig:violins}. First, higher eccentricity does not always lead to better constraints on modified dispersion effects, despite the fact that it improves the bounds on the dimensionless coupling parameter $\bar{\alpha}_{0}$ (see Fig.~\ref{fig:fisher}). The reason for this is entirely due to the divergence of the peak harmonic number $k_{\rm max}$ as $e\rightarrow 1$. 

Second, under optimal conditions, i.e. high SNR, lack of glitches, optimal template searches, etc., the bounds obtained from the eccentric GW bursts studied here can be stronger than the current bounds obtained from the GWTC-3 catalog. Both the GWTC-3 constraints and the results of Fig.~\ref{fig:violins} \&~\ref{fig:ensemble} are summarized in Table~\ref{tab:summary} for convenience. This is despite the fact that the burst signals considered here do not constitute full inspiral waveforms. The improvement appears to be due to the excitation of additional waveform harmonics by non-zero orbital eccentricity. This is not always true, however. For example, for massive gravitons, the projected constraints with both sets of binary masses considered here is worse than the combined constraint from GWTC-3.

Further, from Eq.~\eqref{eq:disp-bound}, one might expect that the bound improves by considering events at higher luminosity distances $D_{L}$. However, this is not necessarily true. The uncertainties in Fisher analysis are known to scale inversely with SNR $\rho$. Since $\rho = \sqrt{(h|h)}$, then $\rho \sim D_{L}^{-1}$, and Eq.~\eqref{eq:disp-bound} is, approximately, independent of the luminosity distance. We say here ``approximately'' because the proportionality coefficient between $\sigma_{\bar{\alpha}_{0}}$ and $\rho$ can depend weakly on the luminosity distance through covariances with other parameters. Further, in this initial study we have neglected cosmological redshift effects, which, if included in this model, would cause the distance appearing in the definition of $\bar{\alpha}$ in Eq.~\eqref{eq:alpha_bar} to be different from the luminosity distance $D_{L}$, which still appears in the waveform amplitude through $\tilde{h}_{0}$ (see Ref.~\cite{Mirshekari:2011yq} for further details in the quasi-circular limit). This will introduce a cosmological redshift $z$ dependent factor into Eqs.~\eqref{eq:alpha_bar} and \eqref{eq:disp-bound} which, in turn, introduces weak dependence on the luminosity distance provided $z\lesssim1$. So in general, we do not expect the projected bounds to change significantly when considering more distant LIGO sources.

One may also be concerned with what happens to parameter uncertainties when considering next-generation GW detectors, such as Einstein Telescope (ET)~\cite{Hild:2008ng,Hild:2010id} and Cosmic Explorer (CE)~\cite{Evans:2023euw}. For both of these detectors, the main effect of relevance to this analysis is an increase in SNR due to the detectors being more sensitive. A reasonable estimate is that these detectors would increase the SNR by an order of magnitude compared to LIGO~\cite{ET:2025xjr, Reitze:2019iox}. Since uncertainties in Fisher analyses scale inversely with SNR, i.e. $\sigma_{a} \propto \rho^{-1}$, this implies that uncertainties will become tighter by approximately an order of magnitude. We say approximately here as there are some subtleties due to the increased low-frequency sensitivity of ET and CE, but the general effect on parameter uncertainties are still controlled by the above SNR argument.

Lastly, it is interesting to compare our projected constraints to those obtained from other observations and experiments beyond GW observations. The most up-to-date bounds on Lorentz invariance violations (LIVs) across all sectors can be found at~\cite{Kostelecky:2008ts}. Of particular interest are the electromagnetic bounds from astrophysical observations, of which the most stringent come from afterglow of GRB 221009A~\cite{LHAASO:2024lub,Vasileiou:2013vra,Yang:2023kjq,Xi:2025ruv}. Constraints from those observations place bounds on LIVs at $\sim10E_{P}$ for $a=3$ and $10^{-8}E_{P}$ for $a=4$, where $E_{P}$ is the Planck energy. These bounds are significantly more stringent than current GW bounds, as well as those forecasted here, owing to the significantly higher energy (or alternatively, higher frequency) of the gamma-ray radiation. Note that a similar phenomena explains the results we found here: eccentric GW bursts probe higher energy scales than quasi-circular binaries in the inspiral phase~\cite{Loutrel:2014vja}, thus leading to slightly stronger constraints for $a\ge 3$.

Regardless of the subtleties of the analysis herein, the final results show a promising future for testing GW dispersion effects once high eccentricity sources are definitively detected. Due to this, it may be worth considering tests beyond the standard parameterized/phenomenological ones considered here and throughout the literature. If GR is modified at sufficiently high energies, then the parameterized dispersion relation in Eq.~\eqref{eq:disp-relation} arises as the low-energy/EFT expansions of whatever the correct high-energy gravitational theory is. Non-linear effects become more important at higher energy scales, and eccentric sources can probe scales higher than those for quasi-circular sources~\cite{Loutrel:2014vja}. In fact, many proposed quantum gravity scenarios and theories introduce non-linear modified dispersion relations, such as logarithmic (causal set theory \& asymptotic safety), exponential (infinite-derivative gravity), and trigonometric (polymer quantization)~\cite{PerezTeruel:2025njj}. Such modification will likely introduce a richer modulation structure to the eccentric bursts beyond the simply Bessel structure found herein. Further, drawing from solid-state physics for some examples, if spacetime is discrete, then a plausible non-linear effect that may arise is the existence of band gaps in the GW energy/frequency spectrum. Band gaps are known to exist in the propagation of electrons in metals~\cite{2008PhRvB..77p5135J,2021PhRvX..11d1010L,Punk:2015fha}, as well as the propagation of light in photonic materials~\cite{Johri:2007,2017OptMa..72..403B,Maagt:2004}. Other examples include scenarios where the dispersion effects are not just dependent on the wave number as in Eq.~\eqref{eq:disp-relation}, but also intensity dependent~\cite{2021SciA....7.3695K,GAETA2005258,Chang:2023,2018RSPSA.47480173H}. Such effects could be interesting avenues for future research.

%%%%%%%%%%%%%%%%%%%%%%%%%%%%%%%%%%%%%%%%%%%%%%%%%%%%%%%%%%%%%%%%%%%%%%%%%%%%%%%%%%%%%%%%%%%%%%%%%%%%%
\section*{Acknowledgments}
%%%%%%%%%%%%%%%%%%%%%%%%%%%%%%%%%%%%%%%%%%%%%%%%%%%%%%%%%%%%%%%%%%%%%%%%%%%%%%%%%%%%%%%%%%%%%%%%%%%%%
We thank Leo Stein for discussions.
A.B. is supported by the International REU Site for Gravitational Physics under NSF grants PHY-2348913 and PHY-1950830.  %\dg{are these for the IREU or something else? We should mention the IREU program somehow}
N.L. and D.G. are supported by 
ERC Starting Grant No.~945155--GWmining, 
Cariplo Foundation Grant No.~2021-0555, 
MUR PRIN Grant No.~2022-Z9X4XS, 
Italian-French University (UIF/UFI) Grant No.~2025-C3-386,
MUR Grant ``Progetto Dipartimenti di Eccellenza 2023-2027'' (BiCoQ),
and the ICSC National Research Centre funded by NextGenerationEU. 
D.G. is supported by MSCA Fellowship No. 101064542--StochRewind, 
MSCA Fellowship No. 101149270--ProtoBH, 
MUR Young Researchers Grant No. SOE2024-0000125,
Computational work was performed at CINECA with allocations 
through INFN and the University of Milano-Bicocca, and at NVIDIA with allocations through the Academic Grant program.

%%%%%%%%%%%%%%%%%%%%%%%%%%%%%%%%%%%%%%%%%%%%%%%%%%%%%%%%%%%%%%%%%%%%%%%%%%%%%%%%%%%%%%%%%%%%%%%%%%%%%
\appendix
%%%%%%%%%%%%%%%%%%%%%%%%%%%%%%%%%%%%%%%%%%%%%%%%%%%%%%%%%%%%%%%%%%%%%%%%%%%%%%%%%%%%%%%%%%%%%%%%%%%%%
\section{Revisiting frequency-domain effective fly-by waveforms}
\label{app:efb-new}
%%%%%%%%%%%%%%%%%%%%%%%%%%%%%%%%%%%%%%%%%%%%%%%%%%%%%%%%%%%%%%%%%%%

The EFB approach seeks to accurately model the GW burst emission from highly eccentric binaries. While a time-domain approach showed strong agreement with numerical leading-PN-order waveforms~\cite{Loutrel:2019kky}, frequency-domain waveforms have stumbled on numerical computation issues~\cite{Loutrel:2019kky}, complicated analytical computations~\cite{Loutrel:2020jfx}, and limitations of suitable approximations~\cite{Loutrel:2023rsl}. We here revisit the original, leading-PN-order, frequency-domain EFB waveform from~\cite{Loutrel:2019kky}, and provide a significant simplification for the construction of current and future EFB waveform templates.

At leading PN order, the two-body dynamics reduce to Keplerian orbits perturbed by 2.5PN order radiation reaction, which constitues the quadrupole approximation of Peters \& Mathews~\cite{Peters:1963ux,Peters:1964zz}. Under the assumption of adiabatic evolution of the orbital elements, the eccentricity $e$, semilatus rectum $p$, and mean anomaly $\ell$ evolve according to~\cite{Loutrel:2019kky}:
\begin{align}
    \label{eq:eoft}
    e(t) &= e_{i} - \frac{304}{15} \eta e_{i} \left(\frac{m}{p_{i}}\right)^{5/2} \left(1 + \frac{121}{304} e_{i}^{2}\right) \ell(t)\,,
    \\
    \label{eq:poft}
    p(t) &= p_{i} \left[1 - \frac{64}{5} \eta \left(\frac{m}{p_{i}}\right)^{5/2} \left(1 + \frac{7}{8} e_{i}^{2}\right) \ell(t)\right]\,,
    \\
    \label{eq:loft}
    \ell(t) &= \frac{n_{i}}{2\pi F_{\rm rr}} \left\{ \exp\left[2\pi F_{\rm rr} \left(t - t_{p}\right) \right] - 1\right\}\,,
\end{align}
where $[e_{i}, p_{i}]$ are the values at pericenter, and
\begin{align}
    n_{i} &= m^{1/2} \left(\frac{1-e_{i}^{2}}{p_{i}}\right)^{3/2}\,,
    \\
    F_{\rm rr} &= \frac{96}{10\pi} \frac{\eta}{m} \left(\frac{m}{p_{i}}\right)^{4} (1-e_{i}^{2})^{1/2} \left(1 + \frac{73}{24} e_{i}^{2} + \frac{37}{96} e_{i}^{4} \right)\,.
\end{align}
The quadrupole order waveform polarizations are then given by Eqs.~\eqref{eq:time_polarizations}, and, when combined with Eqs.~\eqref{eq:eoft}-\eqref{eq:loft}, are valid for $\ell \in [-\pi,\pi]$. To obtain the time-domain EFB-T waveforms, one applies a resummation procedure originally developed in Refs.~\cite{Loutrel:2016cdw, Forseth:2015oua} directly to these waveform polarizations.

However, Ref.~\cite{Loutrel:2019kky} showed that the Fourier transform of the time-domain waveforms can be obtained analytically by application of the SPA, before performing the resummation procedure. After transforming the waveforms into the form of Eq.~\eqref{eq:time_polarizations} and defining $\epsilon = 1 - e^{2}$, application of the SPA gives~\cite{Loutrel:2019kky}:
\begin{align}
    \label{eq:efb-f}
    \tilde{h}_{+,\times}(f) &= -\frac{m^{2}\eta}{D_{L} F_{\rm rr}} \sum_{k=1}^{\infty}\left(\frac{\epsilon_{k,-}^{\star}}{p_{k,-}^{\star}}\right) \frac{\left[H_{+,\times}^{(k)}(e_{k,-}^{\star})\right]^{\dagger}}{\sqrt{2\pi \chi}} e^{i\Psi_{\star}(k, f)},
\end{align}
where $\chi = f/F_{\rm rr}$, $\chi_{\rm orb} = n_{i}/2\pi F_{\rm rr}$, $\Psi_{\star}$ is the waveform's stationary phase
\begin{equation}
    \label{eq:Psi_star}
    \Psi_{\star}(k, f) = k\chi_{\rm orb} - \chi \left[1 +  \ln\left(\frac{k \chi_{\rm orb}}{\chi}\right)\right] - \frac{\pi}{4} + 2\pi f t_{p}\,,
\end{equation}
and $[e_{k,-}^{\star}, p_{k,-}^{\star}, \epsilon_{k,-}^{\star}]$ correspond to the time evolving orbital elements evaluated at the stationary point
\begin{equation}
    t_{k,-}^{\star} = t_{p} + \frac{1}{2\pi F_{\rm rr}} \ln \left(\frac{2\pi f}{k n_{i}}\right)\,.
\end{equation}
With application of the resummation procedure from Refs.~\cite{Loutrel:2016cdw,Forseth:2015oua}, the above waveform can be resummed to obtain the EFB-F waveform which has known complications when one attempts to numerically evaluate the model~\cite{Loutrel:2019kky}, and analytic simplifications to speed up the waveform evaluation are complicated~\cite{Loutrel:2020jfx}.

Instead, we apply a slightly different resummation procedure here. After converting the sum in Eq.~\eqref{eq:efb-f} to an integral over $k$, the integral takes the standard form of a generalized Fourier integral and we apply the SPA again to the stationary phase $\Psi_{\star}(k,f)$. The new stationary point is given by $k_{\star} = \chi/\chi_{\rm orb} = 2\pi f /n_{i}$, and the EFB waveform reduces to
\begin{equation}
    \tilde{h}_{+,\times}(f) = -\tilde{h}_{i} {\cal{H}}_{+,\times}(f; e_{i}) e^{2\pi i f t_{p,i}}\,,
\end{equation}
where $\tilde{h}_{i} = (2\pi/D_{L}) \left({\cal{M}}^{5}/n_{i}\right)^{1/3}$ with ${\cal{M}} = m \eta^{3/5}$ the chirp mass, and the Fourier amplitude is given by
\begin{align}
    {\cal{H}}_{+,\times}(f;e_{i}) = \lim_{k\rightarrow 2\pi f/n_{i}} \left[\Theta(k-1) H_{+,\times}^{(k)}(e_{k,-}^{\star})\right]^{\dagger}\,,
\end{align}
with $\Theta(x)$ the Heaviside step function, which accounts for the fact that the original sum over $k$ begins at $k=1$. It is worth noting that, after both applications of the SPA (first over time, then over $k$ for resummation),
\begin{equation}
    \lim_{k\rightarrow 2\pi f/n_{i}} [e_{k,-}^{\star}, p_{k,-}^{\star}, \epsilon_{k,-}^{\star}] = [e_{i}, p_{i}, \epsilon_{i}],
\end{equation}
and thus,
\begin{equation}
    {\cal{H}}_{+,\times}(f; e_{i}) = \lim_{k\rightarrow 2\pi f/n_{i}} \left[\Theta(k-1) H_{+,\times}^{(k)}(e_{i})\right]^{\dagger}\,,
\end{equation}
which is simply the time-domain harmonic coefficient evaluated at continuous frequency $f$ instead of the discrete harmonic index $k$. This should not be surprising since adiabatic radiation reaction, by its nature, should have little effect on the dynamics of the binary over a single orbit.

As a last point, the amplitude functions $H_{+,\times}^{(k)}(e)$ can be obtained from Eqs.~\eqref{eq:Spk}-\eqref{eq:Ckc} and the relationship 
\begin{equation}
H^{(k)}_{+,\times} = (1/2)(C^{(k)}_{+,\times} - i S^{(k)}_{+,\times})\,.
\end{equation}
These are then dependent on the Bessel function $J_{k}(k e)$ and its derivative $J'_{k}(ke)$. However, $k=2\pi f/n(e,p)$ and when performing the Fisher analysis in Sec.~\ref{sec:params}, one would have to take a derivative of these Bessel functions with respect to their order, not just their argument. To address this, we rely on the uniform asymptotic expansion of Bessel functions of the form $J_{k}(ke)$, specifically~\cite{1965hmfw.book.....A}
\begin{align}
    J_{k}(ke) &\sim \frac{1}{\pi} \sqrt{\frac{2}{3}} \left(\frac{\zeta^{3}}{1-e^{2}}\right)^{1/4} K_{1/3}\left(\frac{2}{3} k \zeta^{3/2} \right) + {\cal{O}}(k^{-1})\,,
    \\
    J'_{k}(ke) &\!\sim\! \frac{1}{e \pi} \sqrt{\frac{2}{3}} \left[\zeta^{3} (1\!-\!e^{2}) \right]^{1/4} \!K_{2/3}\left(\frac{2}{3} k \zeta^{3/2} \!\right) \!+\! {\cal{O}}(k^{-1}),
\end{align}
where $K_{j}(x)$ are modified Bessel functions of the second kind, and
\begin{equation}
    \zeta = \left\{\frac{3}{2} \left[\ln\left(\frac{1+\sqrt{1-e^{2}}}{e^{2}} \right) - \sqrt{1-e^{2}} \right] \right\}^{2/3}\,.
\end{equation}
This ensures that the waveform derivative needed for the Fisher analysis are purely analytical and closed-form.

%\section*{ORCID iDs}
%Nicholas Loutrel \orcidlink{0000-0002-1597-3281} \href{https://orcid.org/0000-0002-1597-3281}{https://orcid.org/0000-0002-1597-3281} \\
%Ava Bailey \orcidlink{0009-0002-0639-906X} \href{https://orcid.org/0009-0002-0639-906X}{https://orcid.org/0009-0002-0639-906X} \\
%Davide Gerosa \orcidlink{0000-0002-0933-3579} \href{https://orcid.org/0000-0002-0933-3579}{https://orcid.org/0000-0002-0933-3579} \\

\section*{References}
\bibliographystyle{iopart-num}
\bibliography{dispersion}

\end{document}